# Single-gate electro-optic beam switching metasurfaces

Sangjun Han[1,†], Jinseok Kong[1,†], Junho Choi[2,3], Won Chegal[2], and Min Seok Jang[1,*]

[*] jang.minseok@kaist.ac.kr

[†] Equally contributed authors

[1] School of Electrical Engineering, Korea Advanced Institute of Science and Technology, Daejeon 34141, Republic of Korea

[2] Strategic Technology Research Institute, Korea Research Institute of Standards and Science, Daejeon 34113, Republic of Korea

[3] Department of Physics, Kyung Hee University, Seoul 02447, Republic of Korea

**Abstract**

Electro-optic active metasurfaces have attracted attention due to their ability to electronically control optical wavefront with unprecedented spatiotemporal resolutions. In most studies, such devices require gate arrays composed of a large number of independently-controllable local gate electrodes that address local scattering response of individual metaatoms. Although this approach in principle enables arbitrary wavefront control, the complicated driving mechanism and low optical efficiency have been hindering its practical applications. In this work, we demonstrate an active beam switching device that provides high directivity, uniform efficiency across diffraction orders, and a wide field of view while operating with only a single-gate bias. Experimentally, the metasurface achieves 57° of active beam switching from the 0th to the -1st order diffraction, with efficiencies of 0.084 and 0.078 and directivities of 0.765 and 0.836, respectively. Furthermore, an analytical framework using nonlocal quasinormal mode expansion provides deeper insight into the operating mechanism of active beam switching. Finally, we discuss the performance limitations of this design platform and provide insights into potential improvements.

# Introduction

Dynamic control of optical beam direction is an emerging technology in a wide range of applications including light detection and ranging (LiDAR), freespace optical communication, laser display, and laser machining. Conventional beam control methods that rely on mechanically moving parts[1] have often suffered from their bulky size and reduced durability. MEMS-based beam control devices combining an actuator and a microscanner and a flash LiDAR based on a diffraction element have been adopted to solve this problem[2]. However, these approaches still suffer from low durability and significant power consumption.

Active metasurfaces provide a promising route to overcome these challenges as they enable a precise control of the wavefront of light with unprecedented spatiotemporal resolutions[3-8]. Active beam switching metasurfaces have been implemented utilizing various active tuning mechanisms including mechanical[9-14], thermal[15-18], and electric[19-21] methods. Among these tuning mechanisms, electrically tunable beam switching method based on electro-optic materials such as indium tin oxide[7,22,23], metallic polymer[24] or transition metal dichalcogenides[25] is attracting attention as it offers a small device footprint, reduced power consumption, minimal heat generation, and improved frame rate over other tuning mechanisms[1,26,27]. Injecting carriers into the electro-optic materials alters their refractive indices modulating the optical response of the metasurface. To overcome small electro-optic index change and maximize the optical modulation of the device, most electro-optic metasurfaces leverage optical resonances to enhance interaction between light and matter[28-30]. In recent years, electro-optic metasurfaces have been demonstrated to have active full-$2\pi$ phase modulation with a nearly constant amplitude by employing two or more tuning parameters per metaatom[6,22] or avoided crossing of two resonances[31] to break the strong correlation between the phase and amplitude response of a resonance. To implement a desired spatial wavefront, individual metaatoms need to be independently gated based on their location in a metasurface with a large number of local gate electrodes[7,22].

Although these approaches in principle enable arbitrary wavefront manipulation, the devices based on local resonance control of metaatoms require complex circuit drivers to individually address the scattering response of each metaatom. This complex driving mechanism can cause potential malfunctions and hinders miniaturization. Furthermore, due to their large losses caused by strong light-matter interaction, these approaches tend to exhibit reduced efficiencies[22,24,25]. Moreover, large-angle beam deflection, which requires a steep spatial phase ramp, is particularly challenging for the devices using unit-cell design approaches due to the crosstalk between the neighboring metaatoms[32]. Consequently, these devices typically exhibit uneven diffraction efficiency or directivity across multiple diffraction orders. To make beam switching technology more practical, the challenge remains to simplify the driving mechanism while improving the performance.

In this study, we design and demonstrate an efficient electro-optic beam switching device that operates with a single global gate instead of an array of numerous local gate electrodes. This metasurface simultaneously has high directivity, uniform and comparable efficiency to conventional electro-optic beam switching devices, and a high deflection angle while operating with only a single-gate bias in the mid-infrared region. To tackle the structural design challenge to get a high performance even with a single tuning parameter, we apply an inverse design using the genetic algorithm rather than adopting a unit-cell design approach based on a locally periodic approximation (LPA). Quasinormal mode analysis reveals that the device operation is based on the interference between a gate-tunable resonant mode and non-resonant background response. Finally, we show that the same design principle can be applied to achieve high-efficiency multi-level beam switching. Our work constitutes a stepping stone towards reliable dynamic optical beam control.

## Result

### Device geometry and fabrication

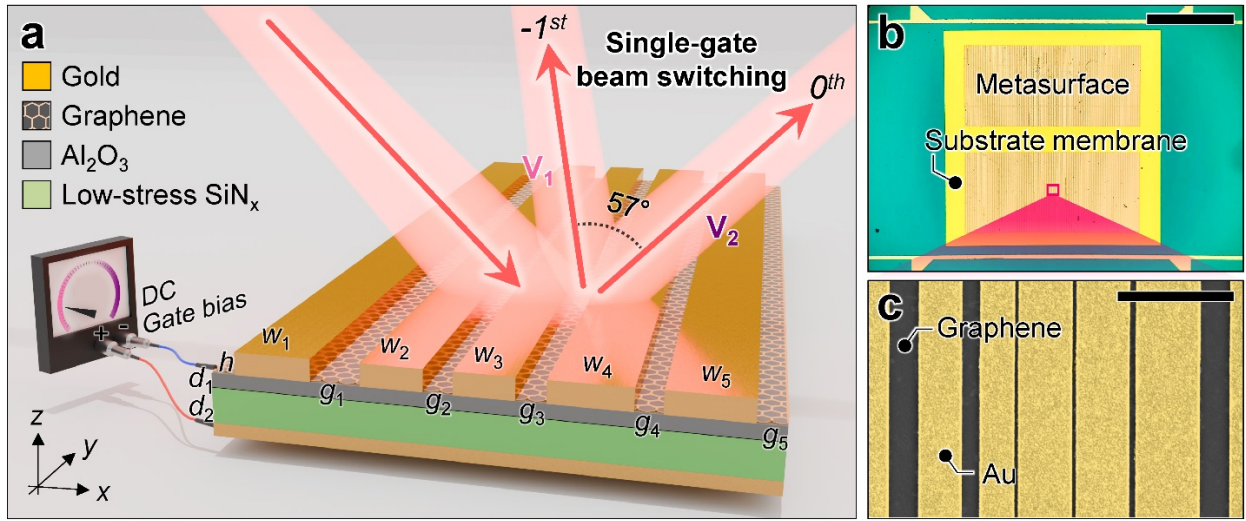

**Fig. 1. Single-gate electro-optic beam switching metasurfaces. a** Schematic of an active beam switching graphene metasurface with gold strip width ($w_1$, $w_2$, $w_3$, $w_4$, $w_5$) = (1135, 914, 1403, 1387, 1671) nm, gap ($g_1$, $g_2$, $g_3$, $g_4$, $g_5$) = (433, 71, 71, 142, 733) nm, gold strip height $h$ = 64 nm, Ti adhesion layer thickness 6 nm, $Al_2O_3$ thickness $d_1$ = 30 nm, and $SiN_x$ thickness $d_2$ = 200 nm. DC gate bias is applied between gold back reflector and monolayer graphene. The operation frequency is $f_0$ = 41.17 THz. **b** Optical microscope top-view image of the fabricated device. Two rectangular regions with deposited gold gratings are located on a substrate membrane which serves as a dielectric layer. To apply a single-gate bias to the graphene layer, two electrode lines are positioned above and below the substrate membrane. The scale bar is 400 μm. **c** Scanning electron microscope (SEM) top-view image of the grating for one period (false colored). Yellow area indicates gold strips and black area indicates gaps where graphene is exposed. The scale bar is 3 μm.

The active beam switching graphene metasurface is designed and fabricated through several intricate steps to achieve desired optical performance in mid-infrared regime. A schematic of the proposed device is illustrated in Fig. 1a. TM-polarized light incident at $\theta_{inc}$ = 45° is reflected by the metasurface with grating
33

period $P$ and diffracted to either the 0th or -1st order channel, depending on the global gate bias applied to the monolayer graphene. We set $P = 7.960$ μm to enable only 0th and -1st order diffraction channels at the operation frequency $f_0 = 41.17$ THz ($\lambda_0 = 7.281$ μm) while suppressing all the higher order diffractions. The -1st order diffraction angle $\theta_{-1} = -11.98°$ is determined by $\sin \theta_{-1} = \sin \theta_{inc} - \lambda_0/P$. The topmost layer of metasurface consists of periodic gold grating, which scatters incident light into 0th or -1st order diffraction channels. A single grating period is composed of five gold strips, each of which has width ($w_i$) and gap ($g_i$) between the neighboring elements. In the gaps between gold strips, monolayer graphenes cover the exposed substrate as shown in Fig. 1a. The substrate is a 200 nm low-stress silicon nitride membrane anchored to the silicon frame with a 30 nm thin film of aluminum oxide deposited on top. This thin alumina layer plays a crucial role for increasing the stability of the device by suppressing gate leakage current[33,34] (see Supplementary Note 1). On the backside of the device, a 70 nm gold layer with a 3 nm Ti adhesion layer serves as a global back gate electrode and also as a back reflector to block transmission channels[28-30].

The key principle of beam switching relies on Fermi level modulation of graphene via the global back gate bias $V_G$ applied between the back reflector and the graphene, which alters the diffraction efficiency of each order[35]. As the Fermi level of graphene switches between the charge neutrality point (CNP) and 0.42 eV, the surface conductivity of the graphene is modified correspondingly[36]. These two Fermi levels are selected to maximize beam switching performance within the stable operation range bounded by the dielectric strength of the gate dielectrics.

Figure 1b shows an overall view of the fabricated device taken under an optical microscope. The low-stress silicon nitride membrane acts as the substrate, supported by a thick silicon frame, facilitating handling of the entire chip. Two straight electrode lines are patterned on either side of the low-stress silicon nitride membrane to apply a gate bias and measure a source-drain current through the graphene channel. The graphene layer extends over the membrane and frame, ensuring uniform gating and modulation across the entire metasurface.

Our metasurface is fabricated through the following steps. With the silicon nitride membrane prepared, a gold back reflector is deposited on the backside using a thermal evaporator. Then, a thin aluminum oxide layer is deposited on the front side of the membrane via atomic layer deposition (ALD). Once the dielectric layers are set up, monolayer graphene is transferred onto the aluminum oxide layer using a wet transfer technique. Electron beam lithography is performed on the graphene layer to specify the location where gold is deposited. To ensure good adhesion of the deposited gold layer, the exposed part of the graphene is etched with an oxygen plasma asher. The device is completed by forming gold gratings through a lift-off process after depositing gold with a thermal evaporator. Detailed fabrication steps are provided in the Methods section. Figure 1c provides a top-view SEM image of the gold grating for one period, confirming that the gold grating and graphene ribbons are clearly defined.

**Single-gate electro-optic beam switching in the mid-infrared regime**

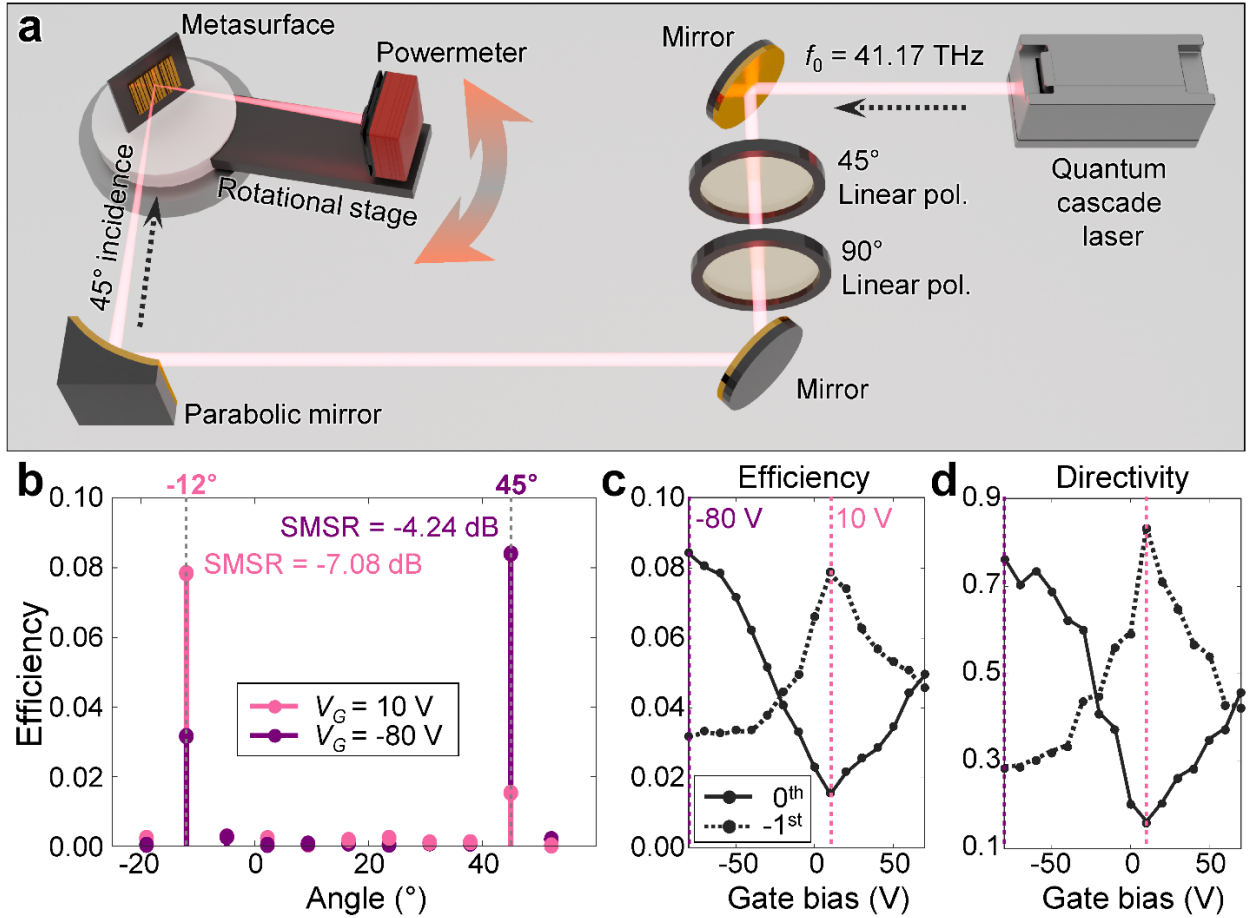

**Fig. 2. Optical setup and measured results. a** Schematic of the optical setup. **b** Angle-resolved far-field pattern measured at the gate bias $V_G = 10$ V and $V_G = -80$ V at the operation frequency $f_0 = 41.17$ THz. At $V_G = 10$ V (pink), highest efficiency is measured at -12° which corresponds to the -1st order diffraction angle, and at $V_G = -80$ V (purple), highest efficiency is measured at 45° which corresponds to the 0th order diffraction angle. **c** Experimentally measured diffraction efficiencies and **d** derived directivities for the 0th and the -1st order diffractions when the gate bias is swept from $V_G = -80$ V to $V_G = 70$ V.

The performance of the single-gate active beam switching is evaluated using an optical setup depicted in Fig. 2a. The experimental setup is built to precisely measure the diffraction efficiency of the reflected light from the metasurface over a wide angular range. Initially, light from the quantum cascade laser is polarized into TM mode by passing through the 45° and 90° polarizers sequentially, and then is focused onto the metasurface using a parabolic mirror. The efficiency of the reflected light is measured with a powermeter mounted on the rotational stage, allowing the observation of diffracted light at various angles. Detailed information of the optical measurement is described in the Methods section.

The angle-resolved far-field pattern is measured at each gate bias $V_G$ at the operation frequency of $f_0 = 41.17$ THz by rotating the stage where the powermeter is mounted, to characterize the quality of the diffracted beam as shown in Fig. 2b. The side mode suppression ratio (SMSR) is also assessed, yielding values of -7.08 dB for the -1st order diffraction at $V_G$ = 10 V and -4.24 dB for the 0th order diffraction at $V_G$ = -80 V, meaning that the light scattered by the metasurface is highly directed into the two diffraction channels with marginal side lobes. The angular interval between measurement points is determined by the angular field-of-view of the powermeter.

The beam switching capability of the fabricated device is further characterized by measuring the optical efficiency and the directivity as a function of the gate bias at the two angular peak positions where the 0th and -1st order diffractions mainly occur as shown in Fig. 2c. The gate bias is swept from -80 V to 70 V. At $V_G$ = 10 V, which corresponds to the CNP, the metasurface exhibits the highest efficiency of 0.078 at -12° (Fig. 2c). This major angle is consistent with the theoretical -1st order diffraction angle, calculated based on the period of the metasurface measured by Atomic Force Microscope (AFM). In contrast, at $V_G$ = -80 V, the major angle shifts to 45°, which is the 0th order diffraction (specular reflection) angle, with the efficiency of 0.084. The directivities of each diffraction order are calculated by taking the ratio of efficiency at each diffraction angle to the total reflected power over the entire measurable angular space (see Supplementary Note 2 for definitions). Figure 2d shows the active tuning of directivity for each diffraction order as a function of the gate bias. At $V_G$ = 10 V, the directivity of the -1st order diffraction is 0.836 and the directivity of the 0th order diffraction at $V_G$ = -80 V is 0.765. Noteworthily, as the gate bias changes from -80 V to 10 V, it can be seen that the directivities of the 0th and -1st order diffractions are gradually crossing, which is the characteristic that could potentially be used as a tunable freespace beam splitter.

**Device design and analysis based on electromagnetic simulation**

The structure parameters and the operation frequency of our metasurface are optimized using a genetic algorithm. A genetic algorithm is an optimization method that mimics the process of natural selection in biological evolution, and it is a classic but well-known for its powerful performance[37,38]. Since it is not gradient-based, it can optimize even discontinuous and complicated figure of merit and has been widely utilized to optimize the structure of metasurfaces[39-42]. To design a metasurface with high beam switching performance, we set the figure of merit to simultaneously maximize efficiency and directivity in the 0th order diffraction at one target Fermi level, and in the -1st order diffraction at the other target Fermi level. Each gene in the gene pool is characterized by the gold strip widths $w_i$, gap sizes $g_i$, the metasurface height $h$, and the target operation frequency $f_0$. Here, $i$ represents the index of the gold strip ($1 \leq i \leq 5$). The optimized structure and the operation frequency are obtained by iterating the selection, cross-over, and mutation processes within the gene pool until the figure of merit is saturated at a certain extremum point[37]. Detailed figure of merit, design parameters, and an optimization flow chart are described in the Supplementary Note 3.

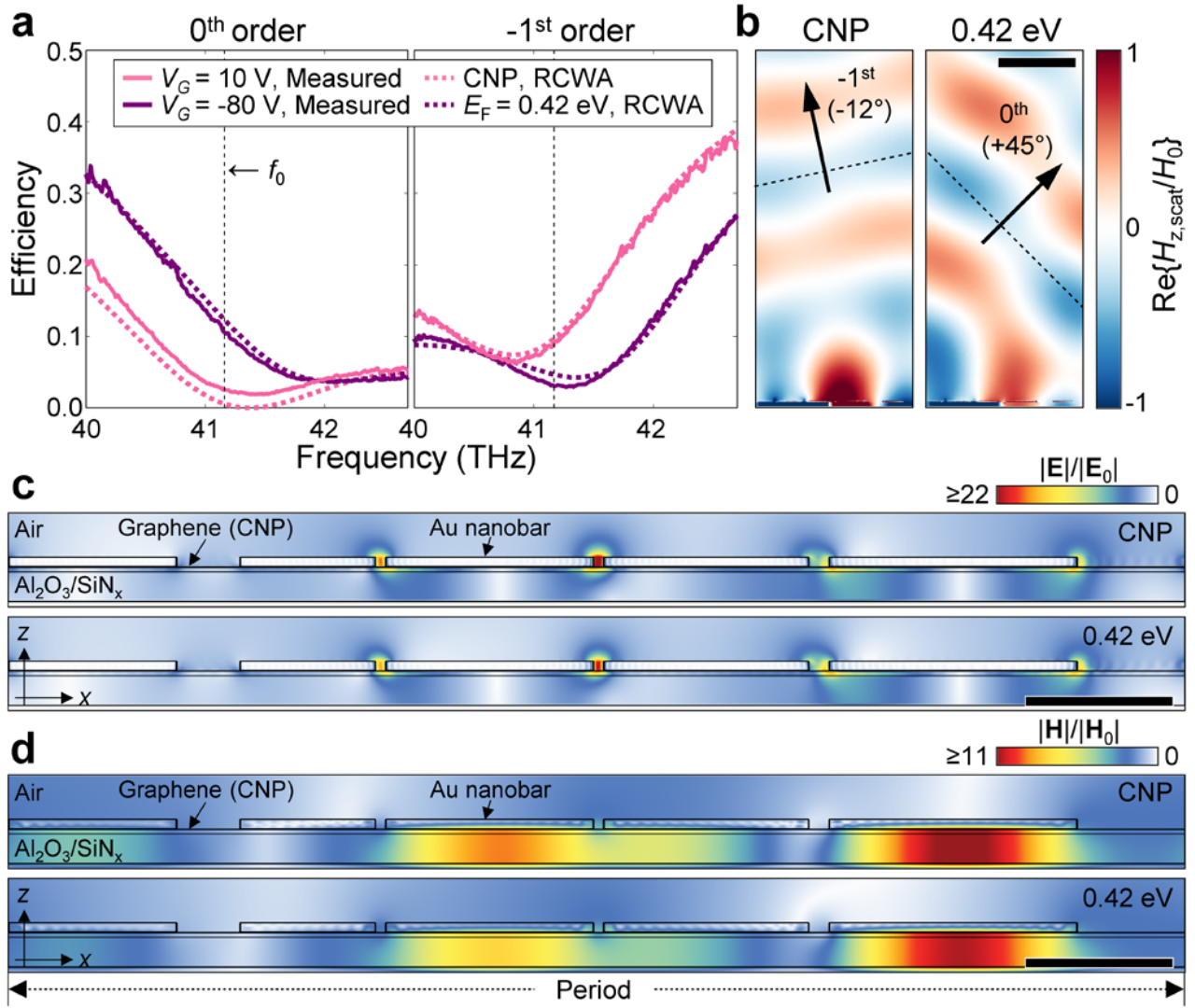

**Fig. 3. Electromagnetic simulation of the single-gate electro-optic beam switching metasurfaces. a** Experimentally measured (solid lines) and electromagnetically simulated (dashed lines) efficiency spectra for the 0th and the -1st order diffractions. **b** Simulated scattered magnetic field profile and **c-d** electromagnetic field intensity profile of the charge neutrality point (CNP) and $E_F = 0.42$ eV at the operation frequency $f_0 = 41.17$ THz. The scale bars are 4 μm in **b** and 1 μm in **c** and **d**.

The actual fabricated device exhibits structure parameters that deviate slightly from the optimal values due to minor fabrication errors. However, our design platform has a high tolerance for these errors, so the figure of merit is not significantly degraded (see Supplementary Note 4). The operation frequency for our experiment is chosen as a slightly shifted value from the design operation frequency based on the highest calculated figure of merit from the efficiency spectrum. This frequency provides an optimal balance between efficiency and directivity, making it ideal to demonstrate the beam switching capability of the metasurface.

To understand the mechanism underlying the active beam switching of the optimized metasurface structure and to validate the experimental results, a detailed numerical analysis is performed using the rigorous coupled wave analysis (RCWA)[43-45]. As shown in Fig. 3a, the efficiency spectra for both the 0th and -1st order diffractions under gate biases $V_G$ of 10 V and -80 V are closely consistent with the experimental data, confirming the reliability of the simulation framework. As the gate bias approaches -80 V from the charge neutrality point of $V_G$ = 10 V, the graphene becomes more conducting (i.e., the real part of graphene permittivity becomes more negative), resulting in a blueshift of diffraction spectra as predicted by the first-order perturbation theory[31,46,47]. Based on a simple parallel plate capacitor model and the measured charge neutrality point from the electrical transport measurement, the gate bias swing from 10 V to -80 V is converted to the graphene Fermi level $E_F$ swing from CNP (0 eV) to 0.42 eV [35], and these values are used in the simulation. A graphene carrier mobility is assumed to be 200 cm²/V·s, which shows the best agreement with the experimental results and consistent with previously observed values at mid-infrared frequencies[30]. This low graphene carrier mobility can be attributed to the impurities induced during the wet-transfer process and $O_2$ plasma ashing used in fabricating graphene ribbons[48], but it is not a major problem as the device is designed to be robust under carrier mobility degradation (see Supplementary Note 4). At the CNP, the 0th order diffraction spectrum is calculated to have a near-zero minimum whereas the measured spectrum has a nonzero finite minimum point. This discrepancy can be attributed to factors such as spatial inhomogeneity of the structure over the illuminated area and the presence of residual charges even in the CNP[49] (see Supplementary Notes 1 and 5).

Figure 3b presents the simulated scattered magnetic field profile at the operation frequency for two Fermi levels (CNP and 0.42 eV). At these Fermi levels, the metasurface exhibits a deflection angle of 57°, shifting from the -1st order diffraction (-12°) to the 0th order diffraction (45°). Both switching states exhibit clear wavefronts with uniform efficiency and high directivity toward the target diffraction order. At the CNP, the wavefront is primarily directed toward the -1st order diffraction with efficiency 0.127 and directivity of 0.993. At $E_F$ = 0.42 eV, the wavefront switches toward the 0th order diffraction with efficiency of 0.130 and directivity of 0.773, showing high performance active beam switching in mid-infrared.

The calculated electromagnetic field intensity profiles around the metasurface for these two switching states are presented in Fig. 3c-d. In both switching states, the electric field is concentrated in the narrower gaps between the gold strips since gold has a high electrical conductivity at mid infrared frequencies and thus the electric potential drop mostly occurs in the gap region where the graphene is exposed. The magnetic field is mostly confined within the dielectric layer between the back reflector and the graphene, spreading over a period of the metasurface. Interestingly, despite the significant change in the far-field wavefront, the intensity distribution of both electric and magnetic fields in the device remains similar regardless of the switching states. From the electromagnetic field distribution that is not tightly bound to the graphene or strongly dependent on $E_F$, it is evident that the operation of our device is not relying upon graphene plasmon resonances, which have been utilized in many previous works to modulate mid-infrared light[6,28-30].

It is important to note that the LPA fails to explain the operation of our device, which implies that interactions between gold strips play a pivotal role in device operation. We investigate the phase and amplitude responses of subunits (gold strips and gaps) composing the metasurface with LPA as detailed in Supplementary Note 6. Ideally, a metasurface designed with LPA should exhibit spatially uniform scattering amplitudes and spatially increasing scattering phases[50]. Additionally, in the active metasurface, that spatial gradient of the scattering phase should be tunable with control parameters. However, we find that the reflective phase gradient does not exhibit the mentioned behavior, indicating that our device should be understood beyond LPA.

**Quasinormal mode analysis**

To gain deeper insights on the operation mechanism of the presented device, we perform quasinormal mode (QNM) analysis. A QNM represents the resonant state in the open and non-Hermitian system which decays over time with a complex eigenfrequency[51]. We can understand the overall optical response of the metasurface as an interference between resonant QNMs and non-resonant background response[52-55]. Figure 4a shows the reflection amplitude in the complex frequency plane at the two representative Fermi levels. Around the operation frequency $f_0$, we identify one positively diverging "pole", which is the eigenfrequency of the QNM of the system. As the Fermi level increases, the eigenfrequency of the QNM blueshifts with decreasing graphene permittivity[31,46,47] (see Supplementary Note 7). This tendency is also consistent with the blueshift of the entire spectrum in Fig. 3a, which is the trajectory line along the real frequency axis. At the same graphene Fermi level, the eigenfrequency of the QNM is the same regardless of the diffraction order. However, it contributes differently to the 0th and the -1st order diffraction spectrum along the real frequency axis. The QNM interferes with the non-resonant background response to produce "zeros" in Fig. 4a. Unlike the eigenfrequency or poles, these zeros are located differently across diffraction orders, even at the same Fermi level. In particular, for the 0th order diffraction at the CNP, the zero lies on the real frequency axis, which means nearly vanishing specular reflection. On the other hand, for the -1st order diffraction at the same Fermi level, the zero is far from the real frequency axis, resulting in the baseline of the reflection spectrum larger than 0. This order-specific difference accounts for the high directivity of the wavefront toward the -1st order diffraction at the CNP.

We then decompose the complex diffraction coefficients into the individual contributions of the resonant QNM and non-resonant background response by employing the Reisz projection method[55] as shown in Fig. 4b. As expected, the optical response by the QNM (black solid lines) clearly blueshifts with increasing $|E_F|$ (i.e., increasing carrier concentration). The background response (black dotted lines) presents non-resonant spectral behavior that marginally depends on $E_F$. The small dependence of the background response on $E_F$ is attributed to the additional QNMs located outside of the Reisz projection contour, far from the real operation frequency (see Supplementary Note 7). The complex sum of the QNM and the background response (red solid lines) agrees well with the RCWA-calculated spectrum (black circles), confirming the reliability of the decomposition. The decomposed electromagnetic field profiles of the QNM and the non-resonant background response are plotted in the Supplementary Figs. S13-14. The total electromagnetic field can be reconstructed by algebraically summing the decomposed field profiles. The reconstructed total field shows good agreement with the results in Fig. 3c-d, which are directly calculated utilizing RCWA under oblique incident plane waves.

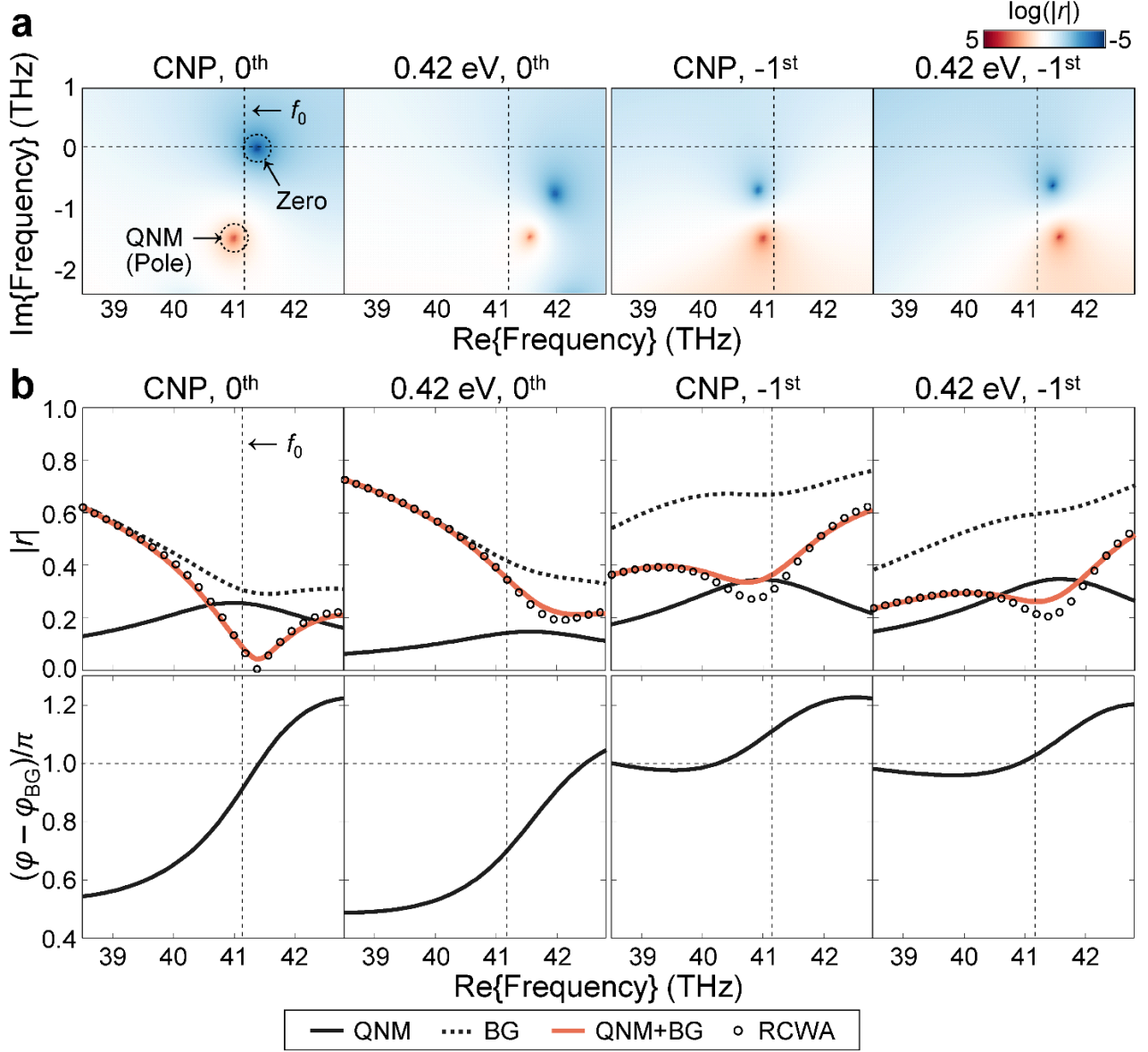

**Fig. 4. Quasinormal mode (QNM) analysis. a** Spectra extended to the complex frequency of the log(|r|) for the 0th and the -1st order diffractions at the charge neutrality point (CNP) and $E_F$ = 0.42 eV. Positively diverging points are the "poles" of the system and negatively diverging points are the "zeros" of the system. **b** Complex diffraction coefficient |r|exp(iφ) spectra for the 0th and the -1st order diffractions at the CNP and $E_F$ = 0.42 eV. In the upper panel, Reflection amplitude |r| spectra are decomposed with resonant QNM and non-resonant background response. Reconstructed reflection amplitude spectra (red solid lines) show good agreement with the electromagnetically simulated spectra (black circles). The lower panel shows the reflection phase difference with the background response φ-$φ_{BG}$ spectra.

The active beam switching behavior of the metasurface can be explained as an interference between the QNM and the background responses. At the CNP, for the 0th order diffraction, they have similar amplitudes and are out of phase, resulting in a nearly perfect destructive interference with a vanishing specular reflection at the operation frequency. In contrast, for the -1st order diffraction, the QNM shows much smaller amplitude compared to the background response. This mismatch in amplitude causes an incomplete cancellation, leading to a significant diffraction amplitude along the -1st order channel. Consequently, the wavefront at the CNP predominantly directed toward the -1st order diffraction. Conversely, at $E_F = 0.42$ eV, the diffraction amplitude difference between the QNM and the background modes remains similar across diffraction orders. However, the phase difference for the -1st order diffraction is much closer to $\pi$ than that of the 0th order diffraction, leading to wavefronts predominantly directed toward the 0th order diffraction.

## Discussion

To further explore the performance potential of the proposed platform, we optimize the structure with a more relaxed fabrication and material quality constraints (see Supplementary Note 8). Here, the minimum feature size of the device structure is set to 20 nm, which can be achieved at the current state-of-art level. For this optimization, we assumed a graphene carrier mobility of 1000 cm²/V·s, which is readily achievable with CVD grown graphene in practical applications[56,57]. Compared to the previous structure designed with tighter fabrication constraints, the newly optimized metasurface with the grating period of 6.097 μm, exhibits a significantly higher theoretical performance at the operation frequency of 44.16 THz as shown in Fig. 5a. For both the 0th and -1st order diffractions, a high efficiency of 0.385 with near-unity directivity of 0.976 is obtained, showing a clear wavefront in Fig. 5a.

Under these relaxed structural constraints, three-level active beam switching, including an additional channel for the 1st order diffraction, is also feasible. To accommodate three diffraction channels, we set the grating period $P = 8.358$ μm and normal beam incidence. At the operation frequency $f_0 = 45.89$ THz ($\lambda_0 = 6.533$ μm), the three diffraction angles are then defined as $|\theta_{-1}| = |\theta_1| = \arcsin(\lambda_0/P)$ for the ±1st orders and $\theta_0 = 0°$ for the 0th order. To enable active switching of the wavefront in three directions, three target Fermi levels are required in total. Here, we set 0.1 eV and 1 eV as the lowest and highest Fermi levels. For intermediate electrical modulation between these values, a Fermi level of 0.7 eV is chosen, where graphene has an average carrier density of the two cases roughly. The optimized metasurface achieves efficiencies of 0.311, 0.148 and 0.148 and directivities of 0.851, 0.852 and 0.886 for the 0th, -1st and 1st order diffractions respectively as shown in Fig. 5b. We anticipate that, with wider design space and more elaborate optimization methodologies, it should be possible to achieve multi-level single-gate beam switching for more than three levels.

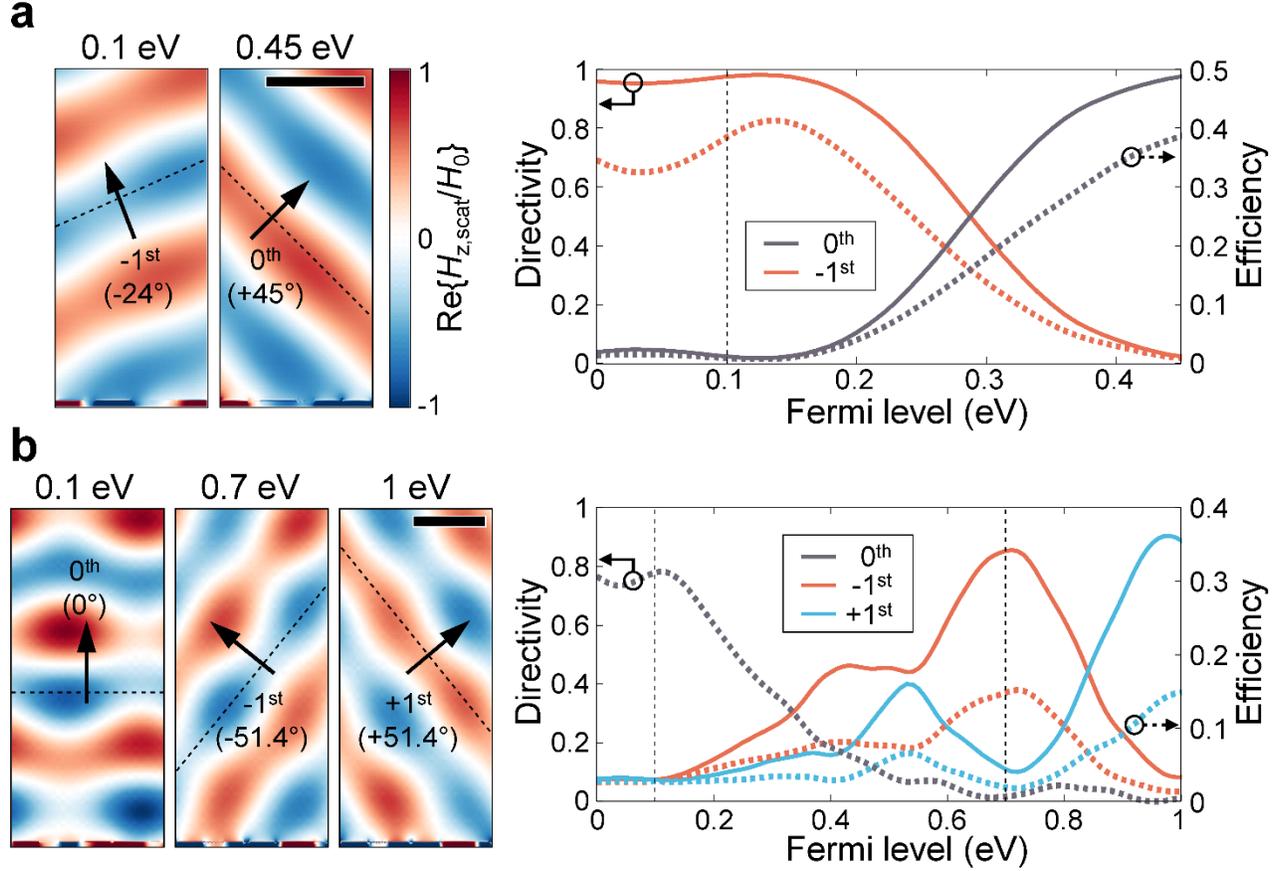

**Fig. 5. Optimization result of the active beam switching metasurfaces under relaxed constraints.** Simulated scattered magnetic field profiles (left panel) and beam switching performance as a function of Fermi level (right panel) **a** for 2-level switching and **b** for 3-level switching. In the right panel, solid lines represent directivity (left axis) and dotted lines represent efficiency (right axis).

Improved device fabrication processes, such as extreme ultraviolet lithography, which enable finer feature sizes with higher precision can further push the boundaries of the functionalities of this metasurface. Furthermore, the development of charge injection mechanisms to increase the Fermi level of the graphene or the use of alternative two-dimensional materials can lead to much higher efficiency and directivity. The operation frequency of this study, the mid-infrared, offers substantial advantages for LiDAR and optical communication. The atmospheric window (8-12 μm) exists in this region and the atmospheric loss rate is significantly lower even under severe weather conditions, compared to the conventionally widely used near-infrared[58]. Thus, it allows for longer detection ranges with lower power consumption regardless of weather conditions, and is safer for human eyes as well as camera sensors. Although this study focuses on the mid-infrared region, the design and analysis principle can be extended to visible and near-infrared frequencies, which are widely adopted for conventional applications.

In conclusion, our work presents an experimental demonstration and comprehensive analysis of the single-gate electro-optic beam switching graphene metasurface, which exhibits a wide deflection angle, high directivity, and uniform efficiency across the diffraction orders originating from Fermi level dependent interference between a resonant QNM and the background modes. Due to the strong interaction between the constituting optical elements, it is difficult to design such a device with a unit-cell design approach relying upon LPA. Instead, we adopt structural optimization approaches based on a genetic algorithm, providing further advancements in the design of active beam switching metasurfaces. This emerging technology has a lot of potential applications, such as LiDAR, optical communication, freespace tunable beam splitters, active control of solar sails, and optical computing. The performance of these metasurfaces is expected to improve with further advancements in material science and device fabrication techniques, leading to even more influential applications and discoveries.

## Methods

### Device fabrication

The metasurfaces were fabricated on a 200 nm-thick low-stress silicon nitride membrane (NX10100D, Norcada). A 70 nm-thick gold back reflector with a 3 nm-thick titanium adhesion layer was deposited on the backside of the membrane using thermal evaporation. On the opposite side of the gold back reflector, a 30 nm-thick aluminum oxide layer was deposited via atomic layer deposition. The top electrode lines were patterned using photolithography with a negative mask and a mask aligner (MJB4 Mask Aligner, SUSS MicroTec), and deposited using thermal evaporation (7 nm titanium/70 nm gold). A CVD grown graphene was directly wet-transferred from the polymethyl methacrylate (PMMA)/Graphene/polymer layers which are purchased from Graphenea Inc. The optimized metasurface structure was then patterned using e-beam lithography with a PMMA resist. The entire size of the metasurface was 818 μm×377 μm which is sufficiently larger than the beam spot size. Utilizing the patterned PMMA layer as a soft etch mask, part of graphene exposed to the air was etched by oxygen plasma asher. The metallic gratings were then formed by life-off of a 64 nm-thick gold layer with a 7 nm titanium adhesion layer deposited by thermal evaporation. The flow chart of the device fabrication steps is provided in the Supplementary Note 9.

**Measurement**

A tunable quantum cascade laser (MIRcat-2400, Daylight Solutions) was employed as a continuous wave monochromatic light source operating in the 6-11 μm frequency range. An infrared step attenuator (102-C, LASNIX) was placed in front of the quantum cascade laser to address the intensity of the laser without distortion of the Gaussian wavefront. The intensity of the light source incident on the metasurface was measured to be 3.37 mW at the operation frequency 41.17 THz. The laser polarization was adjusted to TM mode using two linear polarizers. A gold parabolic mirror focused the laser light to achieve a beam spot diameter of 213 μm at the metasurface. The focused light was incident on the metasurface at an angle of 45º relative to the surface normal vector of the metasurface. The power of the reflected light was measured by a powermeter (PM16-401, Thorlabs) mounted on a plate attached to a high-precision rotation stage that rotates around the sample position. The powermeter has inherent background signal fluctuations over time. The dark state intensities were measured right before and after the metasurface measurement to linearly compensate for this fluctuation (see Supplementary Note 10). The detector size of the powermeter is 10 mm, which can collect the reflected light spreaded over a 7.125° angular region. A Keithley 2400 Sourcemeter was employed to apply the DC gate bias to the metal-dielectric-graphene capacitor and to measure the gate-source and source-drain currents. All measurement was performed in a nitrogen atmosphere to prevent the graphene degradation caused by laser exposure. (see Supplementary Note 11)

**Electromagnetic simulation for structure optimization and analysis**

The device structure parameters were optimized using a genetic algorithm implemented in MATLAB. The figure of merit of individual device structure was calculated using RETICOLO V9, an open MATLAB library for RCWA[43,44]. The detailed definition of the figure of merit, design parameters, and the optimization flowchart are described in the Supplementary Note 5. The reflection coefficient in the complex frequency plane was calculated using S4, an open Python library for RCWA[59]. For QNM expansion (Fig. 4b), RPExpand, an open MATLAB library for Riesz projection expansion of resonance phenomena[55], was combined with S4. To decompose $\mathbf{E}(\mathbf{r},f_0)$ and $\mathbf{H}(\mathbf{r},f_0)$ of the resonant QNM and the non-resonant background response, each field profile was integrated along the contour surrounding the pole and the operation frequency $f_0$ using Cauchy's residue theorem[54]. Details on the integration contour, expansion range and other simulation settings used to draw Fig. 4 are provided in the Supplementary Note 8. In all simulations and optimization processes, the surface conductivity of graphene was calculated based on the random phase approximation[36]. The refractive index of gold was measured through mid-infrared spectroscopic ellipsometry. For titanium and aluminum oxide, data were obtained from Rakić[60] and Kischkat[61], respectively. For low-stress silicon nitride, for which the mid-infrared ellipsometry was not compatible due to its smaller membrane area (1 mm×1 mm) than the beam size of the ellipsometer, the reflection spectrum was measured, and its dispersive relative permittivity was fitted using the Brendel-Bormann model to reconstruct the refractive index. The relative permittivity of all materials was analytically continued for the complex frequency analysis (Supplementary Note 12).

## Author contributions

S.H. and J.K. contributed equally to this work. S.H. and M.S.J. conceived the ideas. S.H. and J.K. conducted optical simulations, fabricated the metasurfaces, and measured the experimental data. J.C. and W.C. measured the refractive index of gold. S.H. conducted the quasinormal mode analysis. J.K. optimized metasurface design parameters. S.H., J.K., and M.S.J. wrote the manuscript. J.C. and W.C. contributed to the discussion and the revision of the manuscript. M.S.J. supervised the project.

# Supplementary Information for

# **Single-gate electro-optic beam switching metasurfaces**

Sangjun Han[1,†], Jinseok Kong[1,†], Junho Choi[2,3], Won Chegal[2], and Min Seok Jang[1,*]

[*] jang.minseok@kaist.ac.kr

[†] Equally contributed authors

[1] School of Electrical Engineering, Korea Advanced Institute of Science and Technology, Daejeon 34141, Republic of Korea

[2] Strategic Technology Research Institute, Korea Research Institute of Standards and Science, Daejeon 34113, Republic of Korea

[3] Department of Physics, Kyung Hee University, Seoul 02447, Republic of Korea

**This file includes:**

Supplementary Notes 1 to 13

1. Electrical properties of the device
2. Efficiency, directivity and SMSR
3. Metasurface design parameter optimization
4. Design parameter tolerance
5. Spatial inhomogeneity within the metasurface
6. Analysis using locally periodic approximation
7. Poles, zeros, and Riesz projection
8. Optimization of a metasurface with relaxed constraints
9. Device fabrication steps
10. Reduction of the background signal fluctuation
11. Effects of a nitrogen atmosphere
12. Material refractive index fitting
13. Angular divergence of the reflected beam

# Supplementary Note 1. Electrical properties of the device

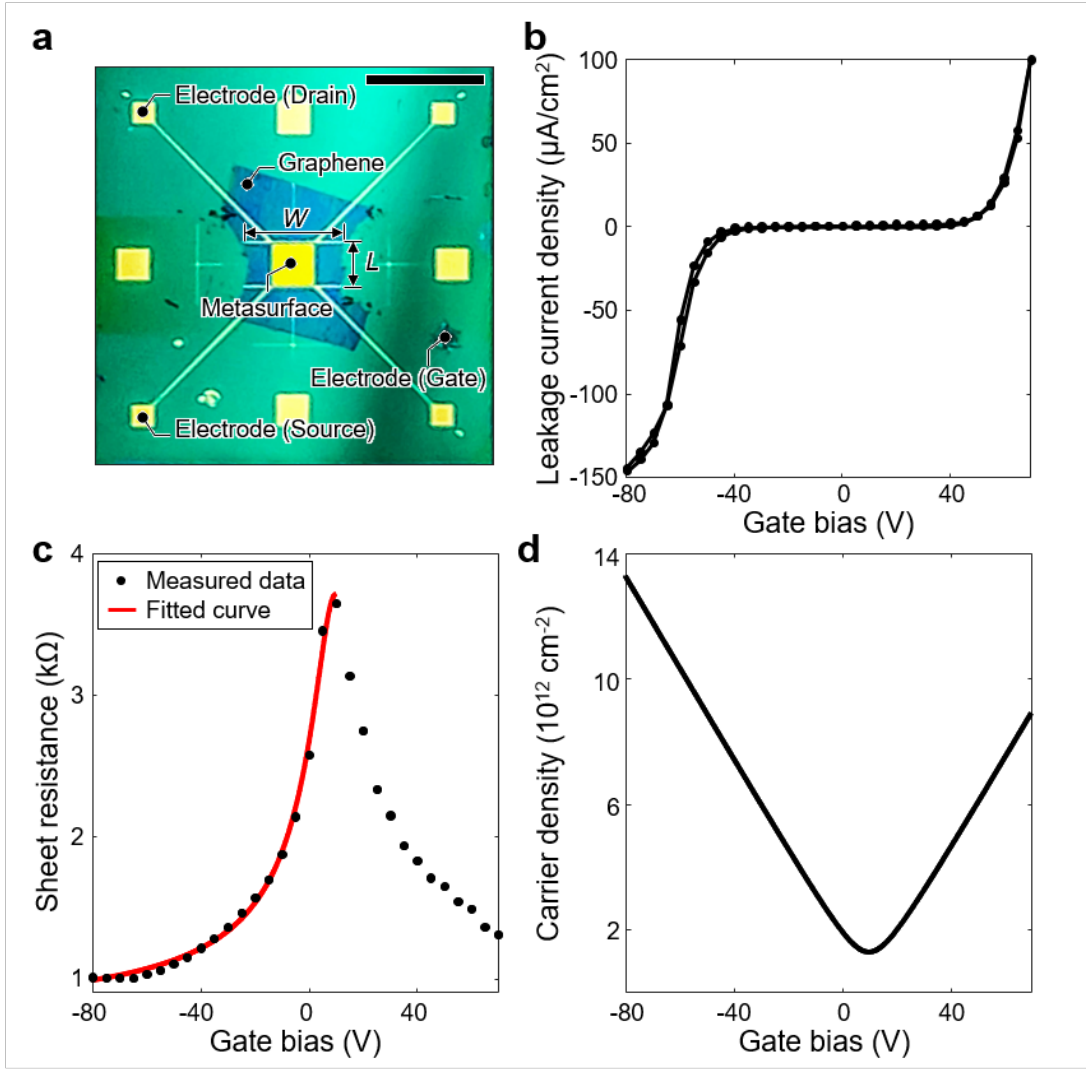

**Fig. S1. Electrical properties of the electro-optic beam switching metasurfaces. a** Top view image of the sample chip. The scale bar is 3 mm. Gate bias $V_G$ dependent **b** leakage gate-source current density, **c** graphene sheet resistance, and **d** calculated carrier density in the graphene. Here, the gate bias at the CNP of this graphene $V_{CNP}$ is 10 V.

## 1.1 Leakage current of the dielectric substrate

To measure the gate-source, source-drain current on this metasurface, electrodes were placed around the metasurface as shown in Fig. S1a. The gate electrode is electrically connected to the back reflector, which is the bottom electrode. The source and drain electrodes are located on the front side of the sample chip and have a mirror symmetric structure. These two electrodes are connected by graphene to form a channel with channel length $L = 1.0$ mm and channel width $W = 2.5$ mm. The top electrode formed by the graphene and the metasurface, the bottom electrode (a back reflector), and the dielectric substrate between them form a metal-dielectric-graphene capacitor. The dielectric substrate is 200 nm of $SiN_x$ and 30 nm of $Al_2O_3$. $SiN_x$ exhibits a significant leakage current at high gate bias $V_G$ and is prone to breakdown, limiting the usable Fermi level range of the device. Depositing 30 nm of $Al_2O_3$ onto $SiN_x$ by atomic layer deposition mitigates this leakage current[1,2]. Fig. S1b shows that this dielectric substrate reliably endures gate bias swing cycles from $V_G = $ -80 V to $V_G = $ 70 V without breakdown.

**1.2 Calculation of the doped carrier density in the graphene**

From the graphene channel formed between the source and drain electrodes, the carrier density doped in the graphene can be deduced. Figure S1c shows the measured DC sheet resistance as a function of gate bias $V_G$. The DC sheet resistance $R_{tot}$ is described by Eq. S1 below:

$$R_{tot} = R_{contact} + R_{channel} = R_{contact} + \frac{L/W}{n_{tot} e \mu}. \tag{S1}$$

Here, $R_{contact}$ is the contact resistance between the electrode and graphene, $\mu$ is the carrier mobility and $n_{tot}$ is the carrier density in the graphene channel region. The $n_{tot}$ is expressed by Eq. S2 below:

$$n_{tot} = \sqrt{n_0^2 + \left(\frac{C|V_{CNP} - V_G|}{e}\right)^2}. \tag{S2}$$

$C$ is the capacitance of the metal-dielectric-graphene capacitor. $n_0$ is the residual charge density present even in the charge neutrality point (CNP)[3]. By combining equations S1 and S2 and fitting the data in Fig. S1c (black dots), the values of $R_{contact}$, $n_0$ and $\mu$ can be determined (red solid line). Figure S1d shows $n_{tot}$ as a function of gate bias $V_G$. Due to the residual charge density $n_0 = 1.263 \times 10^{12}$ cm$^{-2}$, $n_{tot}$ exhibits a larger carrier density than zero even at the CNP ($V_G = 10$ V). Because of this residual charge density, the experimental value measured at the CNP may present discrepancy with the electromagnetic simulation which calculates Fermi level of graphene at the CNP as 0 eV.

## Supplementary Note 2. Efficiency, directivity and SMSR

The main performance metrics used in this study are efficiency, directivity, and side mode suppression ratio (SMSR). The main variables of the metrics are the measurement angle $\theta$ and the applied gate bias $V_G$. The measurement angle $\theta$ is defined as an angle between the normal vector of the metasurface and the vector from the center of the metasurface to the center of the powermeter. Efficiency $E(\theta, V_G)$ is defined as shown in Eq. S3 below:

$$E(\theta, V_G) = \frac{P_{metasurface}(\theta, V_G)}{P_{mirror}(\theta, V_G)}. \tag{S3}$$

Where $P_i(\theta, V_G)$ is the integrated light power reflected from the object $i$ within the angular range of $(\theta - \Delta\theta/2, \theta + \Delta\theta/2)$ at the gate bias $V_G$. Here, the angular field-of-view of the powermeter $\Delta\theta$ is $\Delta\theta = 7.125°$. By definition, efficiency can have a value between 0 and 1. Directivity $D(\theta, V_G)$ is the performance metric which evaluates how well the beam is concentrated in a single direction. It is defined as shown in Eq. S4 below:

$$D(\theta, V_G) = \frac{E(\theta, V_G)}{\sum_\theta E(\theta, V_G)}. \tag{S4}$$

In this paper, directivity is defined as the efficiency at an angle $\theta$ to the sum of the efficiencies measured from all measurable angles. By definition, directivity can have a value between 0 and 1. SMSR($V_G$), a metric similar to directivity and already widely used, is defined as shown in Eq. S5 below:

$$\text{SMSR}(V_G) = 10 \log_{10} \frac{E(\theta_{max,side}, V_G)}{E(\theta_{max,main}, V_G)}. \tag{S5}$$

Here, $\theta_{max,main}$ and $\theta_{max,side}$ are the angles at which the efficiency maximizes in the main and side lobes, respectively. The SMSR can theoretically reach -∞ dB when the side lobes are perfectly suppressed.

# Supplementary Note 3. Metasurface design parameter optimization

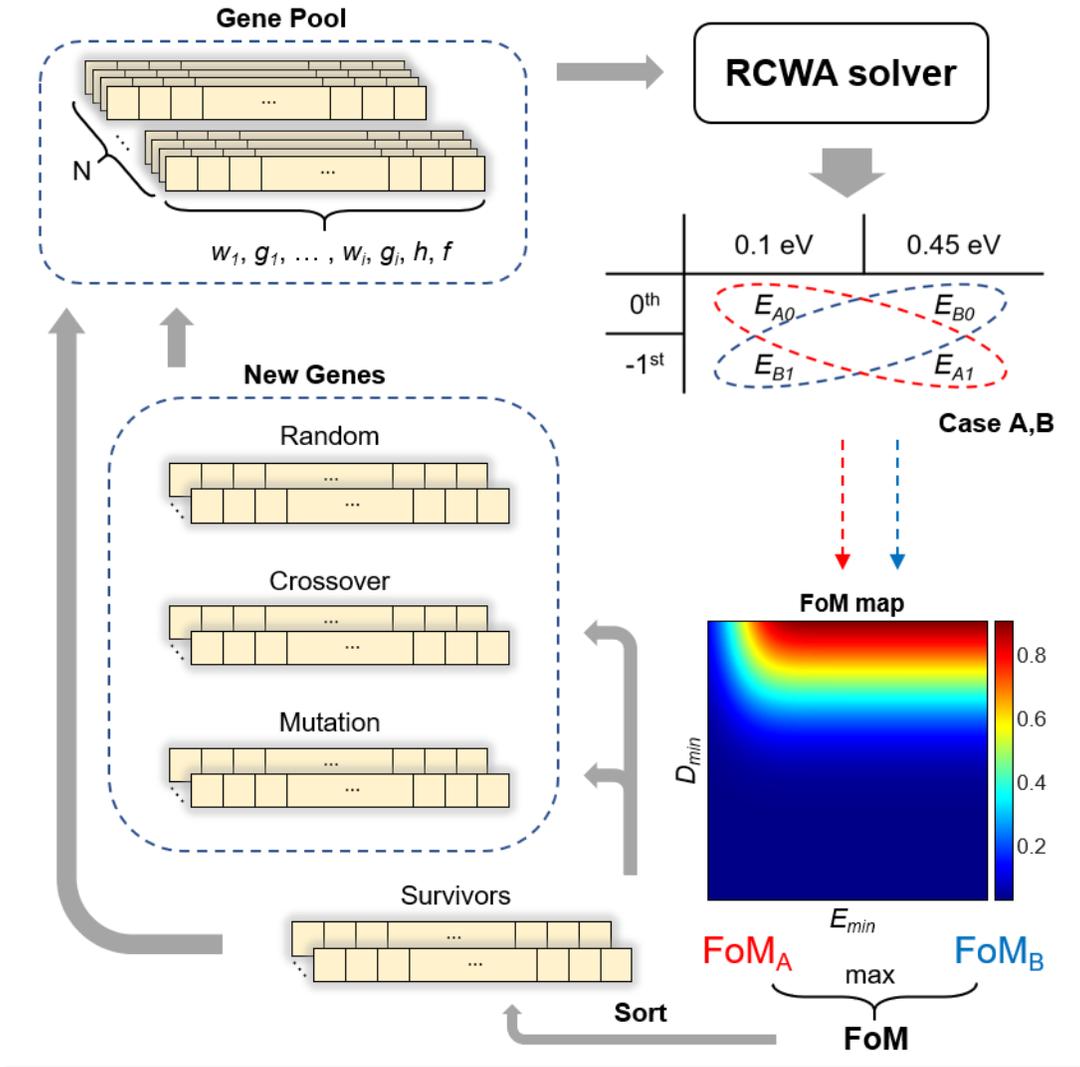

**Fig. S2.** An overview of the optimization flowchart explaining the implementation of the genetic algorithm and the definition of the applied figure of merit (FoM). Case A is indicated in red and case B is indicated in blue.

## 3.1 Parameter setting

The device used in this study was optimized according to the process outlined in Fig. S2. The metasurface is designed with five subunits ($1 \leq i \leq 5$) per period, where each subunit consists of a width and a gap. Therefore, the optimization parameters of the metasurface consist of 12 variables, including five widths ($w_i$), five gaps ($g_i$), the gold strip height ($h$), and the operation frequency ($f_0$). To ensure good adhesion of the gold strip to the dielectric layer, Ti adhesion layer is applied, which has the thickness of $0.1h$, determined by the gold strip height. The lower and upper bounds for each optimization parameter are as follows: 400 nm ≤

$w_i \leq 3000$ nm, $100$ nm $\leq g_i \leq 1200$ nm, $20$ nm $\leq h \leq 70$ nm, $30$ THz $\leq f_0 \leq 50$ THz. The optimization parameters are constrained not only by these bounds, but also by the limitations of the optical setup, which corresponds to the minimum detectable angle $\theta_{min} = -25°$, and the requirement to suppress higher diffraction orders except for the 0th and -1st orders. For a metasurface with period $P$ and incident light at an angle $\theta_{inc} = 45°$ with wavelength $\lambda = c_0/f_0$, where $c_0$ is the speed of light in vacuum, these constraints are expressed as in Eq. S6-7 below:

$$\sin\theta_{min} < \sin\theta_{inc} - \frac{\lambda}{P} < 1, \tag{S6}$$

$$\left|\sin\theta_{inc} - \frac{2\lambda}{P}\right| > 1. \tag{S7}$$

The carrier mobility of the graphene is assumed to be 500 cm$^2$/V·s.

### 3.2 Calculation of FoM

The metasurface used in this study operates based on the working principle that diffraction efficiency of each order changes as the Fermi level of graphene transitions between two values. For the optimization, two specific Fermi levels, $E_{F1} = 0.1$ eV (corresponding to the CNP) and $E_{F2} = 0.45$ eV, are used. Depending on which diffraction order predominantly occurs at each Fermi level, there are two possible cases, A: 0th order diffraction at $E_{F1}$ and -1st order diffraction at $E_{F2}$, and B: -1st order diffraction at $E_{F1}$ and 0th order diffraction at $E_{F2}$. In each case ($x$ = A, B), the figure of merit (FoM$_x$) is calculated using the minimum diffraction efficiency $E_{min,x} = \min(E_{x0}, E_{x1})$ and the minimum directivity $D_{min,x} = \min(D_{x0}, D_{x1})$ of the predominant diffraction orders at the two Fermi levels. The FoM equation is derived by transforming the error function $erf(x) = \frac{2}{\sqrt{\pi}}\int_0^x e^{-t^2}dt$, which is in the form of step function, as shown in Eq. S8 below:

$$FoM_x = \left(erf\left((E_{min,x} - 0.1) \times 8\right) + 1\right) \times \left(erf\left((D_{min,x} - 0.8) \times 4\right) + 1\right)/4. \tag{S8}$$

Here, the values 0.1 and 0.8 represent the target minimum efficiency and the target minimum directivity respectively, as the desired performance for beam switching. The diffraction efficiencies are calculated using RETICOLO V9, an open MATLAB RCWA (Rigorous Coupled-Wave Analysis) library[4,5], with a Fourier order of 175. Between the two calculated FoM values, the larger FoM is selected as the final FoM = max(FoM$_A$, FoM$_B$).

### 3.3 Genetic algorithm

The optimization process is conducted using a genetic algorithm, which treats each set of optimization parameters as a "gene"[6,7]. By altering the composition of the gene pool, which is a set of genes, across generations, the genetic algorithm derives the optimal set of parameters. The gene pool consists of four main categories: survivor genes, random genes, crossover genes, and mutation genes. These genes are generated through the following processes.

- Survivor genes: The top 10% of genes from the previous generation, ranked by their FoM, are selected as survivor genes.
- Random genes: Random genes are generated randomly from the parameter space, constrained by the lower and upper bounds of each parameter. The values of each parameter follow a uniform distribution within the parameter space.
- Crossover genes: Crossover genes are generated from three randomly selected survivor genes from the current generation. For each parameter in a crossover gene, a random selection is made from the corresponding parameters of the three parent genes.
- Mutation genes: Mutation genes are generated by selecting one survivor gene from the current generation as a parent. Each parameter of the mutation gene is determined in one of three ways, with specific probabilities assigned to each method. First, a value is sampled from a normal distribution centered on the parent gene's parameter, with a standard deviation of 5%, which is also bounded by the parameter space (50% probability). Second, a value is directly inherited from the parent gene (37.5% probability). Third, a value is randomly sampled from a uniform distribution within the parameter space (12.5% probability).

Only genes generated by the above method that satisfy Eq. S6-7 are included in the gene pool. Initially, the gene pool starts with 100 random genes. As generations progress, the number of random genes ($N_r$) gradually decreases and approaches zero while the numbers of crossover genes ($N_c$) and mutation genes ($N_m$) increases. By the end of optimization, mutation genes constitute the majority. The detailed changes in the numbers of each type of gene, including the number of survivor genes ($N_s$), across generations are as follows:

$$\text{generation} = 0: (N_s, N_r, N_c, N_m) = (0, 100, 0, 0),$$
$$1 \leq \text{generation} < 10: (N_s, N_r, N_c, N_m) = (10, 54, 18, 18),$$
$$11 \leq \text{generation} < 20: (N_s, N_r, N_c, N_m) = (10, 0, 45, 45),$$
$$21 \leq \text{generation} < 40: (N_s, N_r, N_c, N_m) = (10, 0, 30, 60),$$
$$41 \leq \text{generation}: (N_s, N_r, N_c, N_m) = (10, 0, 18, 72).$$

After 64 generations, the genetic algorithm ends as FoM converges. The convergence of FoM over generations with the changes in $E_{min}$ and $D_{min}$ can be observed in Fig. S3a. The final optimal structure derived from the optimization process has gold strip width $w_i$ = (1176, 914, 1491, 1436, 1735) nm, gap $g_i$ = (464, 100, 100, 185, 1735) nm, gold strip height $h$ = 64 nm, and operation frequency $f_0$ = 40.49 THz. Theoretically, this structure has FoM of 0.689, exhibiting an efficiency of 0.216 and a directivity of 0.927 in the -1st order diffraction at $E_{F1}$ = 0.1 eV, and an efficiency of 0.217 and a directivity of 0.925 in the 0th order diffraction at $E_{F2}$ = 0.45 eV. The Fourier order convergence of the calculated FoM, $E_{min}$, and $D_{min}$ for both the optimal structure and the fabricated structure (as presented in Fig. 1a) in RCWA simulations can be checked in Fig. S3b-c.

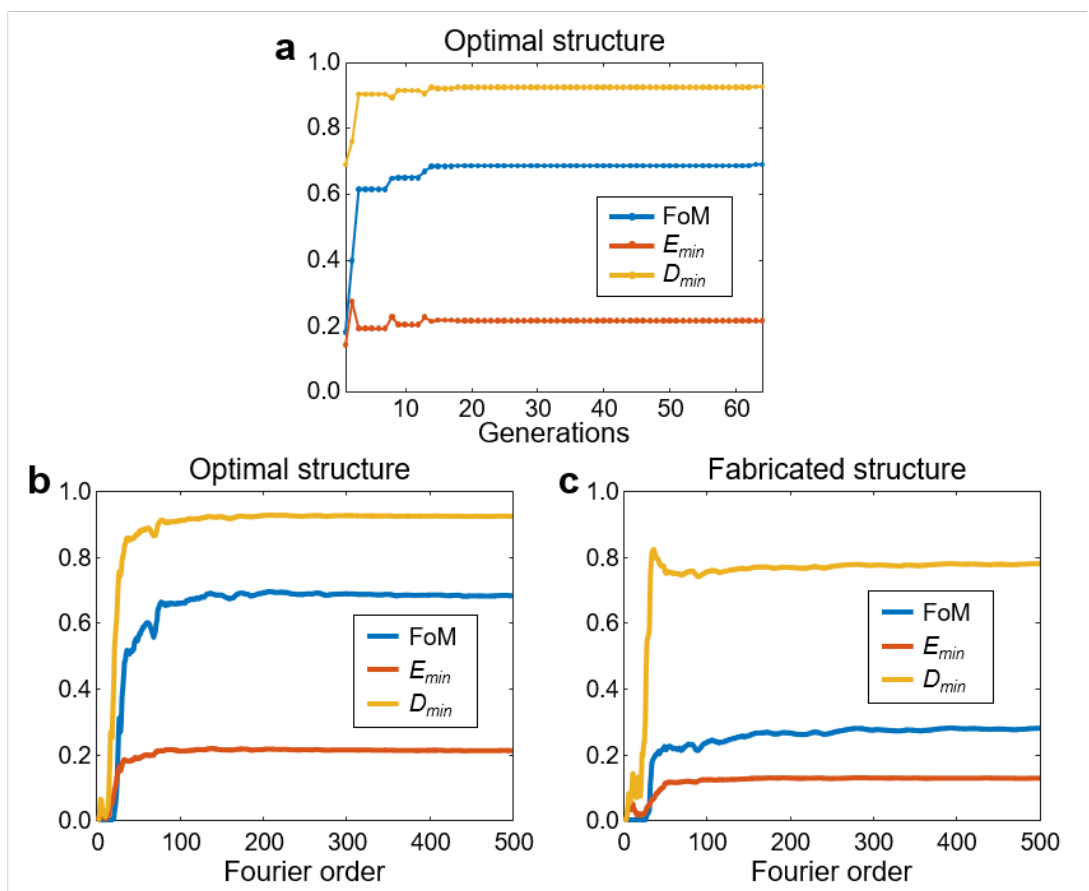

**Fig. S3. The convergence of simulation results with increasing generation and Fourier order. a** Maximum FoM (blue) and corresponding $E_{min}$ (red) and $D_{min}$ (yellow) per generation in the genetic algorithm. **b** Fourier order convergence analysis of FoM, $E_{min}$, and $D_{min}$ for the optimal structure and **c** the fabricated structure.

# Supplementary Note 4. Design parameter tolerance

## 4.1 Optimal structure

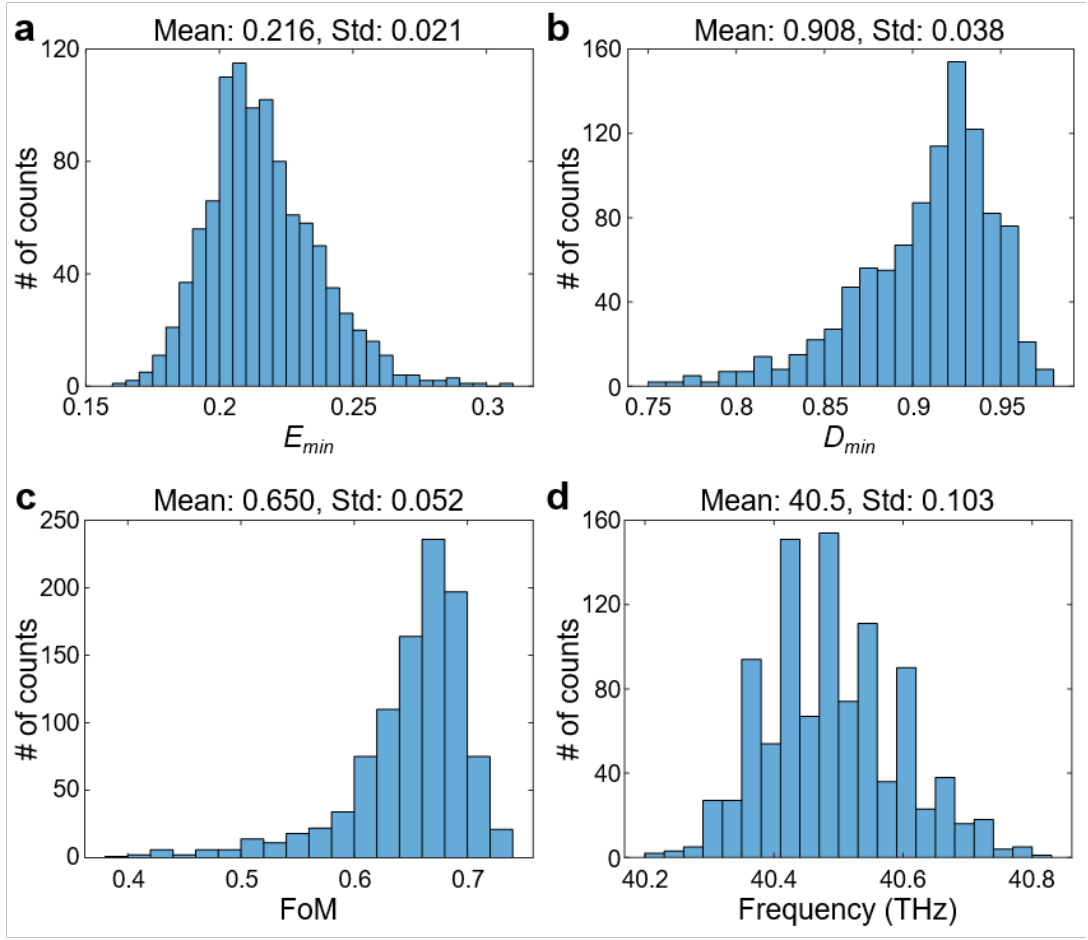

**Fig. S4. Results of the Monte Carlo simulation for the optimal structure. a** Distribution of $E_{min}$ values, with a mean of 0.216 and a standard deviation of 0.021. **b** Distribution of $D_{min}$ values, with a mean of 0.908 and a standard deviation of 0.038. **c** Distribution of FoM values, with a mean of 0.650 and a standard deviation of 0.052. **d** Distribution of optimal operation frequencies, with a mean of 40.5 THz and a standard deviation of 0.103 THz.

To evaluate the impact of structural errors from the fabrication process on beam switching performance, the structural tolerance of the optimal structure is assessed using a Monte Carlo simulation[8]. Considering the characteristics of the electron beam lithography (EBL) process, where structural errors primarily occur, fabrication errors ($\Delta_i$) are applied to each subunit by changing the values of $w_i$ and $g_i$ to $w_i+\Delta_i$ and $g_i-\Delta_i$ respectively, while maintaining the total length of each subunit ($w_i + g_i$) and gold strip height. Here, $Δ_i$ follows a normal distribution with a mean of 0 nm and a standard deviation of 15 nm. The $E_{min}$, $D_{min}$, and FoM of the errored structures are investigated at their optimal operation frequencies, by sweeping the operation frequency within the range of 40.49 THz±1 THz. Figure S4a-d shows the results of the Monte Carlo simulation for 1000 cases. Most of the FoM values of the errored structures are distributed above 0.6, with a mean of 0.650 and a standard deviation of 0.052, which is comparable to the FoM of the optimal structure, 0.689. Additionally, as shown in Fig. S4d, the adjusted operation frequencies are closely distributed around the operation frequency of the optimal structure, 40.49 THz. This theoretical analysis suggests that fabrication errors around 15 nm do not significantly degrade the device's beam switching performance or significantly shift the operation frequency.

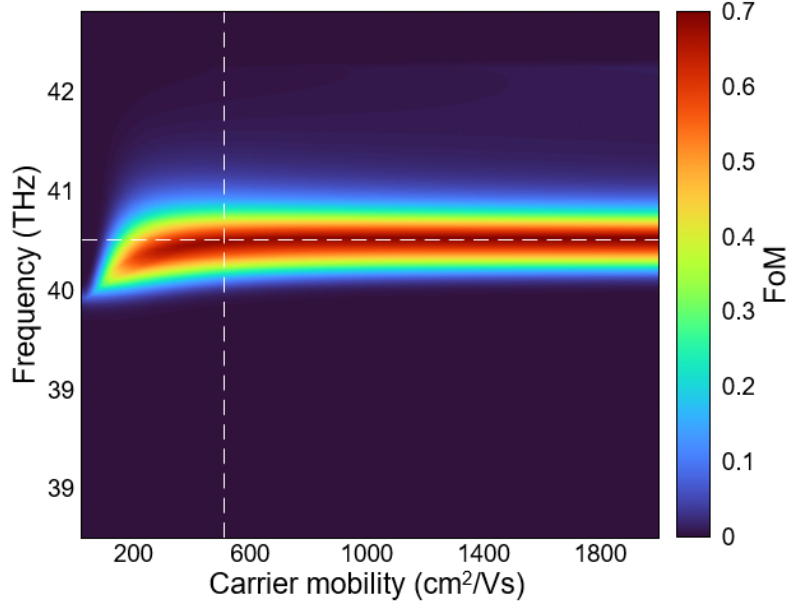

**Fig. S5.** 2D spectrum of the FoM as a function of operation frequency and graphene carrier mobility for the optimal structure. The location of optimal operation frequency $f_0 = 40.49$ THz at the carrier mobility of 500 cm$^2$/V·s is indicated by white dashed lines.

In addition to the structural parameters, we examine how the carrier mobility of graphene affects beam switching performance. The mobility tolerance is evaluated by plotting a 2D spectrum of the FoM as a function of operation frequency for carrier mobility values ranging from 10 to 2000 cm$^2$/V·s, as shown in Fig. S5. When the carrier mobility is 2000 cm$^2$/V·s, a maximum FoM of 0.708 is achieved, which is only marginally higher than the FoM of 0.689 obtained with the carrier mobility of 500 cm$^2$/V·s used in the optimization process. Furthermore, even when the carrier mobility decreases to 150 cm$^2$/V·s, the FoM remains at 0.512, indicating that the beam switching performance is not significantly degraded by reduced carrier mobility and is robust to the change of carrier mobility.

## 4.2 Fabricated structure

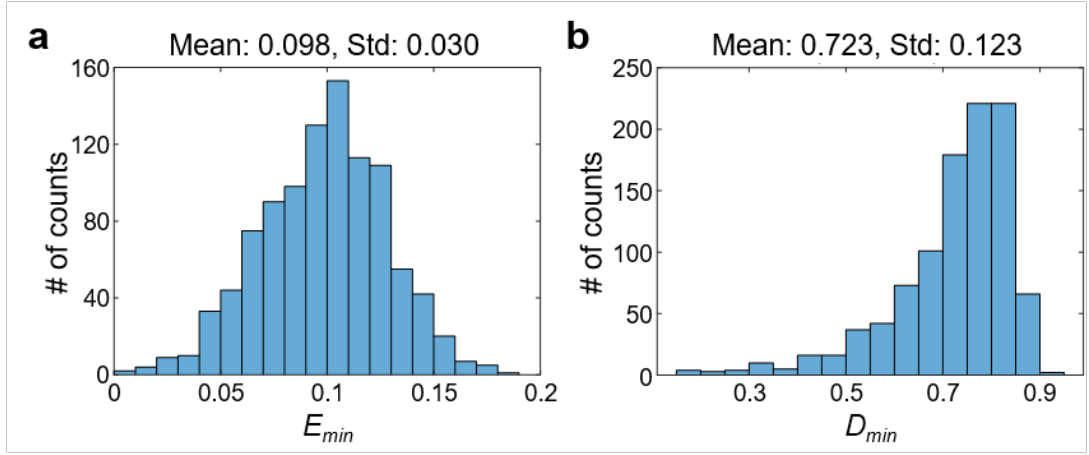

**Fig. S6. Results of the Monte Carlo simulation for the fabricated structure. a** Distribution of $E_{min}$ values, with a mean of 0.098 and a standard deviation of 0.03. **b** Distribution of $D_{min}$ values, with a mean of 0.723 and a standard deviation of 0.123.

The fabricated metasurface does not maintain a uniform pattern across the entire area due to the proximity effect occurring during the EBL process, leading to spatial inhomogeneity[9] (see Supplementary Note 5). As a result, each structural parameter from different partitions may have a different optimal operation frequency. However, since a single operation frequency is used during measurement, it is important to investigate how spatial inhomogeneity affects the beam switching performance of the fabricated structure. This effect is introduced into the Monte Carlo simulation using the same method applied in the structural tolerance analysis for the optimal structure, except that the operation frequency is fixed to a single value. The Monte Carlo simulation is conducted for 1000 cases with $f_0 = 41.17$ THz, graphene carrier mobility of 200 cm$^2$/V·s, $E_{F1} = 0$ eV, and $E_{F2} = 0.42$ eV, based on the experimentally measured diffraction efficiency spectra. Fabrication errors with a standard deviation of 15 nm are applied. As shown in Fig. S6, the results show $E_{min}$ with a mean of 0.098 and a standard deviation of 0.03, and $D_{min}$ with a mean of 0.723 and a standard deviation of 0.123. These values indicate some degradation compared to the simulation result of the fabricated structure without errors, where $E_{min} = 0.127$ and $D_{min} = 0.773$.

Figure S7a illustrates the mobility tolerance of the fabricated structure, evaluated using the same method as for the optimal structure. At a carrier mobility of 200 cm$^2$/V·s, FoM of 0.272 is achieved, which is significantly lower than the maximum FoM of 0.639 observed at a carrier mobility of 2000 cm$^2$/V·s. The detailed mobility tolerance for diffraction efficiencies and directivities for each diffraction order at the operation frequency of 41.17 THz is shown in Fig. S7b-e. Unlike when the Fermi level is at the charge neutrality point, a carrier mobility of 200 cm$^2$/V·s is insufficient for these values to reach saturation when the Fermi level is 0.42 eV.

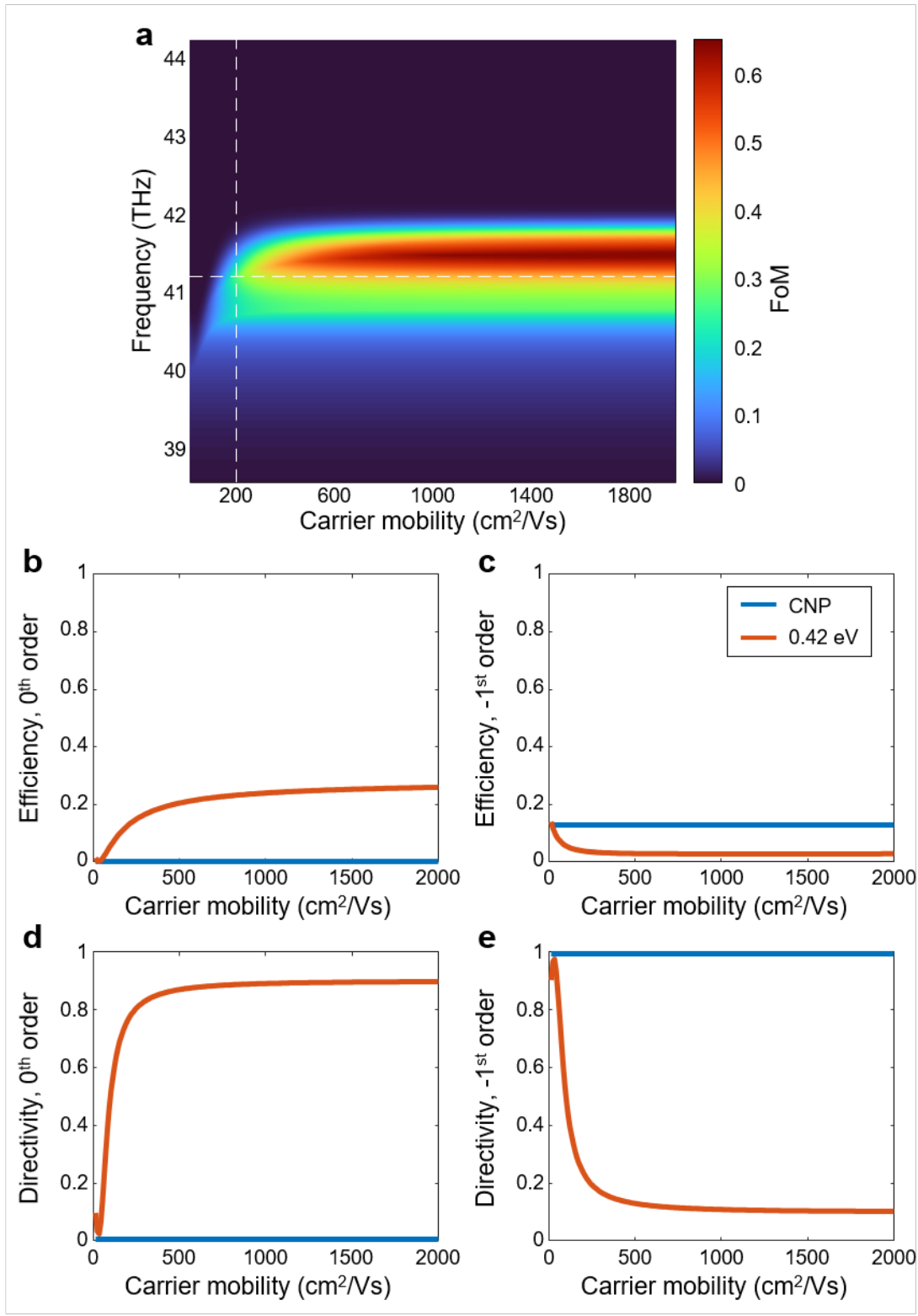

**Fig. S7. Carrier mobility tolerance of the fabricated structure. a** 2D spectrum of the FoM as a function of operation frequency and graphene carrier mobility for the fabricated structure. The location of optimal operation frequency $f_0 = 41.17$ THz at the carrier mobility of 200 cm$^2$/V·s is indicated by white dashed lines. **b** Diffraction efficiencies for the 0th and **c** -1st order diffractions, and **d** directivities for the 0th and **e** -1st order diffractions of the fabricated structure at $f_0 = 41.17$ THz as a function of carrier mobility, at the CNP (blue) and $E_F = 0.42$ eV (red).

These results contrast with the high tolerance observed in the optimal structure, and this discrepancy is related to the incidence angle tolerance. The optimal structure achieves its best performance at an incidence angle of 45°, which is the same value used throughout the simulations. However, as shown in the 2D spectrum in Fig. S8, the optimal incidence angle of the fabricated structure shifts above 45°, indicating that the mobility tolerance analysis of the fabricated structure at a fixed 45° incidence angle may not correspond to that of the optimal structure. Supporting this, the fabricated structure exhibits a high FoM of 0.550 at an incidence angle of 52.5° and an operation frequency of 40.41 THz, even with a low carrier mobility of 200 cm$^2$/V·s. This suggests that at its optimal incidence angle, the fabricated structure is likely to exhibit higher mobility tolerance. Due to limitations of the experimental setup, measurements are not performed at incidence angles other than 45°. If measurements could be taken over a wider range of incidence angles, it is anticipated that the device would show better beam switching performance experimentally.

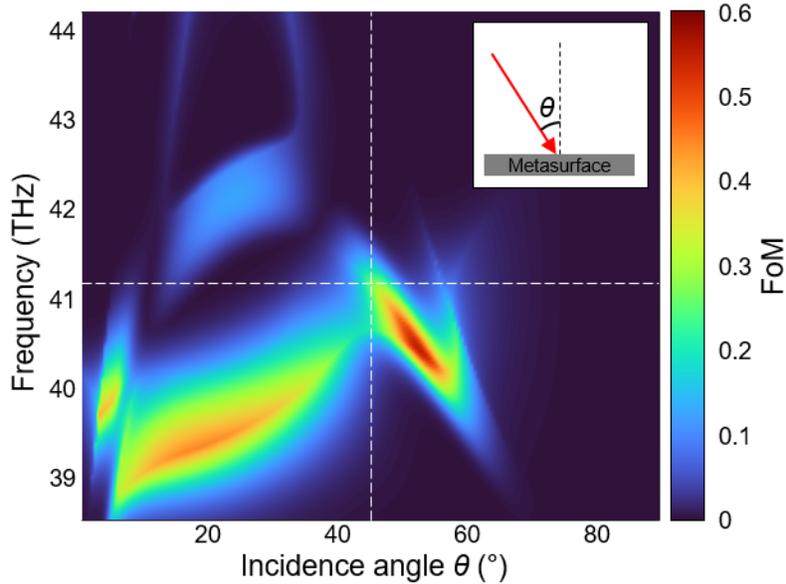

**Fig. S8.** 2D spectrum of the FoM as a function of operation frequency and incidence angle for the fabricated structure at the carrier mobility of 200 cm$^2$/V·s. The location of optimal operation frequency $f_0 = 41.17$ THz at the incidence angle of 45° is indicated by white dashed lines.

# Supplementary Note 5. Spatial inhomogeneity within the metasurface

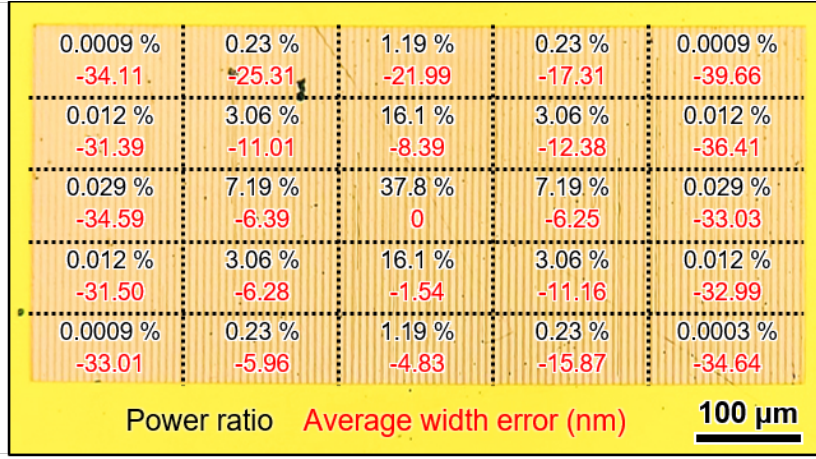

**Fig. S9.** Spatial inhomogeneity distribution of the incident Gaussian beam power and the structure within the metasurface area.

During the metasurface fabrication, a proximity effect can occur during e-beam lithography over a large area. Due to this proximity effect, at the edges of the metasurface, structures can form with different gold strip widths $w_i$ ($1 \leq i \leq 5$) from the center while the length of the subunit $w_i+g_i$ is the same. Also, since the incident laser is a Gaussian beam, there is a difference in illuminated intensity depending on the position of the metasurface. Fig. S9 presents the spatial inhomogeneity of the incident Gaussian beam power and the structure by dividing the entire metasurface area into 5×5 regions. The average width error $\Delta w_{avg}$ (red numbers) is defined as the average of the differences between the gold strip width $w_i$ with that of the center region $w_{i,center}$. (i.e. $\Delta w_{avg} = (\Sigma_{i=1-5} (w_i - w_{i,center}))/5$.) The gold strip widths of each region are measured with a scanning electron microscope. The black numbers are the theoretically calculated ratio of Gaussian beam power, that obliquely illuminates the region with 45º, to the total incident Gaussian beam power. Here, the Gaussian beam diameter is 213 μm. 37.8% of the incident Gaussian beam illuminates the center region. Most of the remaining 62.2% of the beam illuminates the regions that exhibit marginal average width errors. In order to reconstruct the optical response of the corresponding metasurface with electromagnetic simulation very accurately, both spatial inhomogeneity of the Gaussian beam and the metasurface structure might need to be considered.

# Supplementary Note 6. Analysis using locally periodic approximation

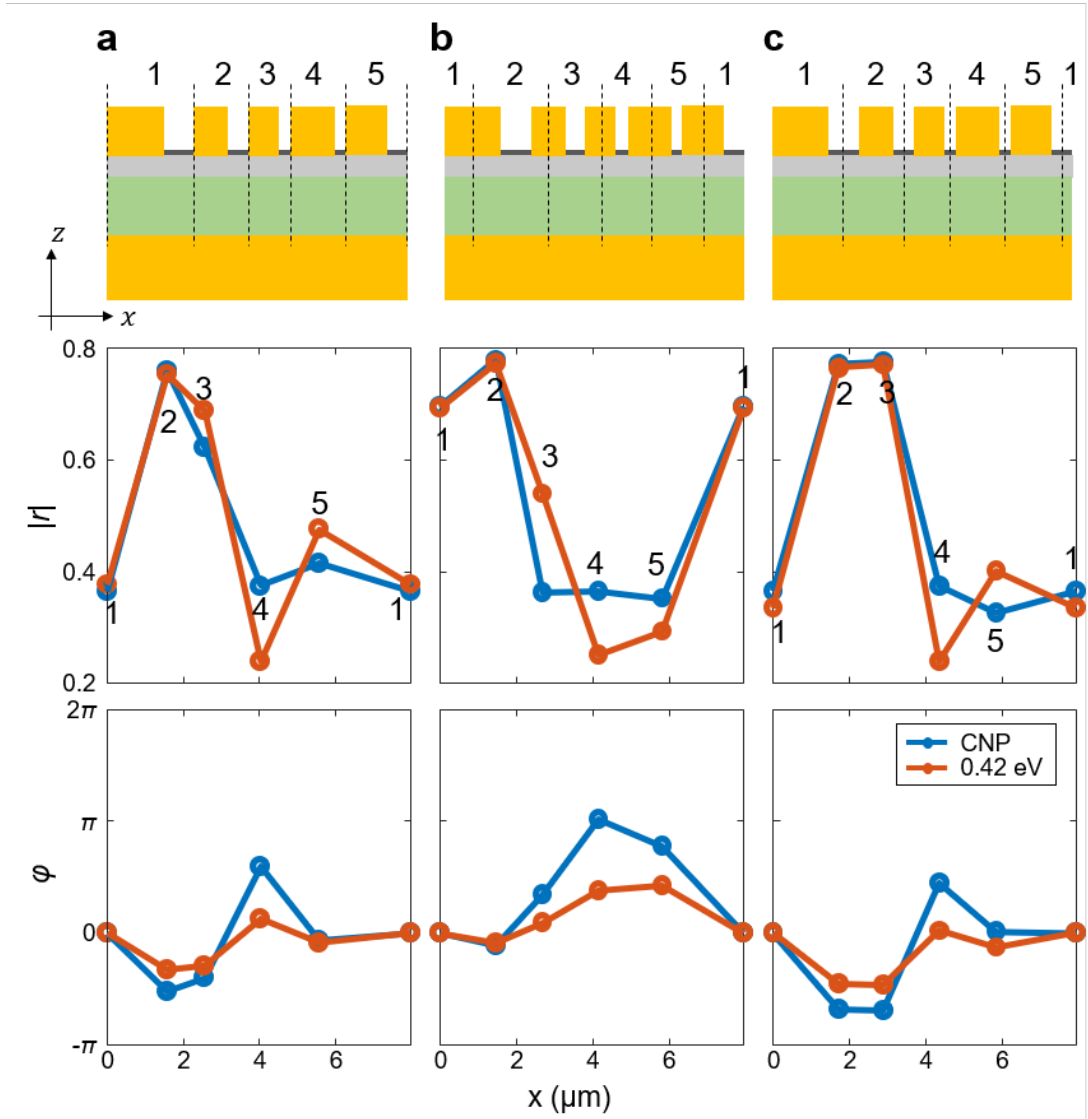

**Fig. S10. Analysis using locally periodic approximation.** Schematics of divided unit-cells (top panel), spatial reflection amplitude distributions (middle panel), and spatial reflection phase distribution (bottom panel) of metasurface divided in three different ways. Metasurfaces are divided into five unit-cells by vertical lines crossing **a** the left side of the gold strip, **b** the center of the gold strip, or **c** the center of the gap.

We analyze the electro-optic beam switching mechanism using a unit-cell method based on the locally periodic approximation (LPA). We divide this metasurface into five subwavelength unit-cells with different gold strip widths and gaps to get the spatial distribution of the local optical response. There are several ways to divide the metasurface grating into unit-cells. This is illustrated in Fig. S10 (top panel). According to LPA, the local optical response of each unit-cell is assumed to be equal to the optical response of a virtual structure of which the unit-cell is spatially repeated infinitely periodically[10]. Figure S10 shows the spatial distribution of the complex reflectivity $|r|\exp(i\varphi)$ within one grating period of the metasurface divided by three different methods. The simulations were performed at a frequency of $f_0 = 41.17$ THz with RETICOLO V9, an open MATLAB RCWA library[4,5]. The electro-optic beam switching behavior of this structure deflected to the -1st order diffraction at the CNP should exhibit a spatially increasing reflection phase distribution from 0 to $2\pi$ within one grating period and a uniform reflection amplitude distribution[11], but the bottom panel of Fig. S10 shows that none of these methods reproduce this behavior. This is because there is non-negligible interference between neighboring unit cells, and each cannot be considered in isolated metaatoms. Thus, in order to design or analyze our electro-optic beam switching metasurfaces, we should utilize another method beyond the LPA.

## Supplementary Note 7. Poles, zeros, and Riesz projection

**7.1 Fermi level dependency of the position of poles and zeros**

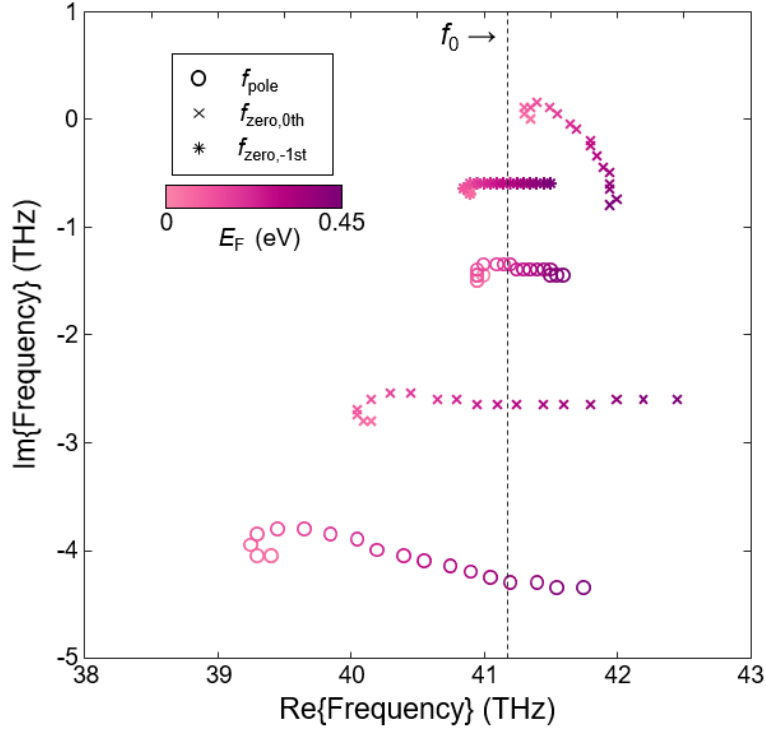

**Fig. S11.** Movement of the pole and zero positions of this metasurface as the graphene Fermi level changes.

Figure S11 shows movements of the QNM pole and the zeros corresponding to the 0th and -1st order diffraction by the graphene Fermi level $|E_F|$. The behavior of these poles can be explained by perturbation theory[12,13]. The resonance frequency shift $\Delta\omega$ can be approximated to the first-order, leading to:

$$\Delta\omega = -\frac{\omega_0}{2}\frac{\int \Delta\varepsilon(\mathbf{r})|\mathbf{E}(\mathbf{r})|^2 dv}{\int \varepsilon(\mathbf{r})|\mathbf{E}(\mathbf{r})|^2 dv}. \tag{S9}$$

Where $\omega_0$ and $\mathbf{E}(\mathbf{r})$ are the resonance frequency of the mode and its electric field, respectively. $\varepsilon(\mathbf{r})$ and $\Delta\varepsilon(\mathbf{r})$ are the permittivity distribution of the materials and its variation by the tuning parameter, respectively. For this metasurface, the tuning parameter is the Fermi level $E_F$ of graphene, and $\Delta\varepsilon(\mathbf{r})$ has a non-zero value only at graphene sites. The volume integral is performed over one grating period of the metasurface. Since the metasurface has a more non-local electric field than a typical graphene metasurface[14-16], the electric field $\mathbf{E}(\mathbf{r})$ is weak enough to apply a first-order perturbation approximation in Eq. S9, and the resonant frequency shift $\Delta\omega$ is roughly proportional to the change in graphene permittivity $\Delta\varepsilon(\mathbf{r})$. As $|E_F|$ increases, the QNM pole first redshifts slightly and then blueshifts as |EF| becomes larger. This is directly related to the change in the real part of the graphene permittivity with $|E_F|$. As $|E_F|$ becomes larger, the graphene becomes more conducting, and the real part of the graphene permittivity becomes more negative ($\Delta\varepsilon(\mathbf{r}) < 0$). Therefore, according to Eq. S9, resonance frequency $\Delta\omega$ blueshifts for large $|E_F|$. And since the position of zeros are determined by the interference between the QNMs and the background response, they show a similar behavior to the QNM poles.

## 7.2 Quasinormal mode analysis with extended imaginary frequencies

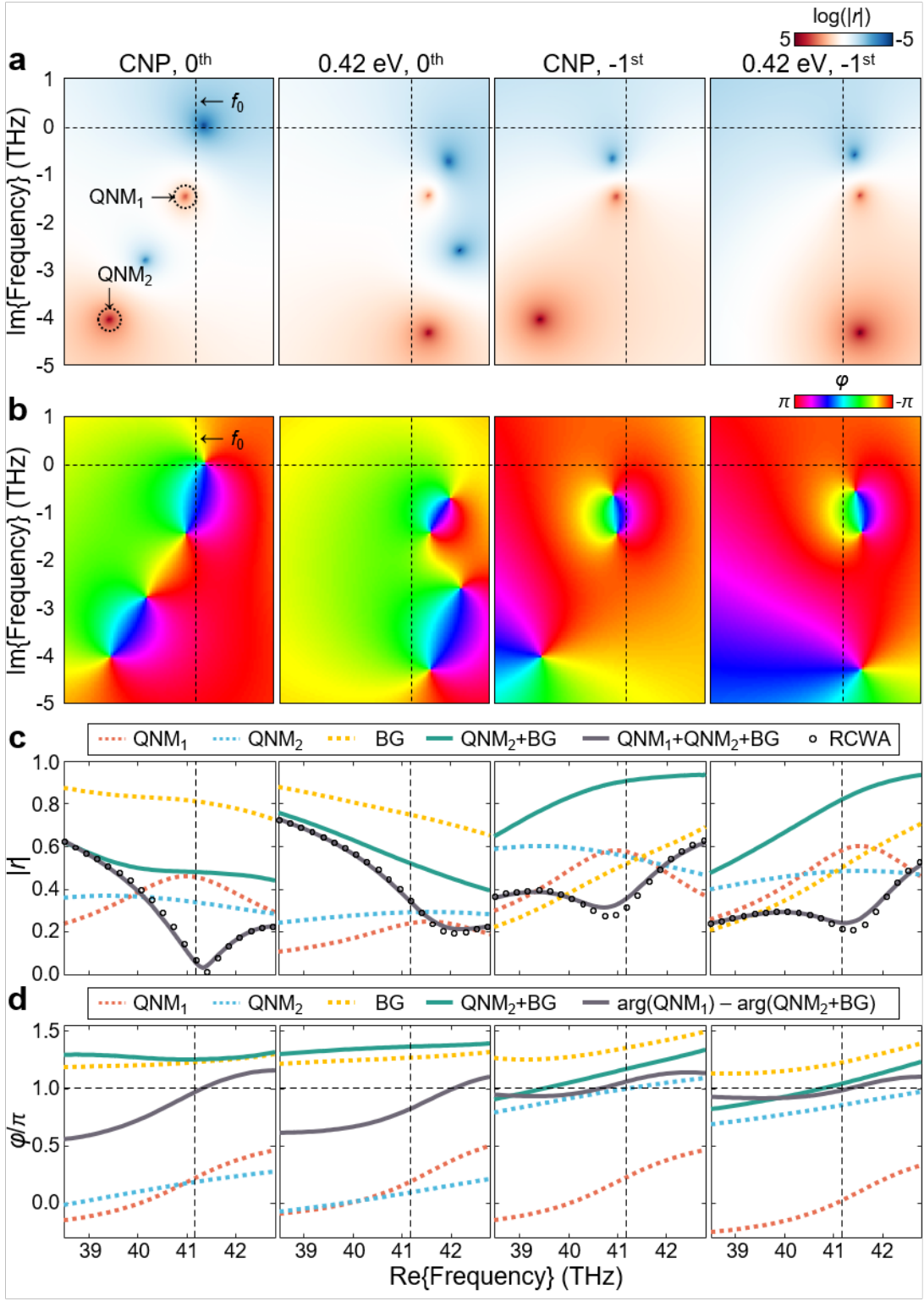

**Fig. S12. Quasinormal mode analysis with extended imaginary frequencies.** Spectra of the complex diffraction coefficient $|r|\exp(i\varphi)$ with extended imaginary frequencies. **a** Logarithm of the amplitude $\log(|r|)$ and **b** phase $\varphi$ are depicted for the 0th and -1st order diffraction at the CNP and $E_F = 0.42$ eV. **c** Amplitude $|r|$

and **d** phase $\varphi/\pi$ spectra decomposed with contributions of the $QNM_1$, $QNM_2$ and the background response (BG) using Riesz projection (dotted lines). In **c**, reconstructed amplitude spectra (gray solid lines) show good agreement with the electromagnetically simulated spectra (black circles). Green solid lines in **c** and gray solid lines in **d** are auxiliary spectra that exhibit similar behavior with the 'background' response spectra in Fig. 4b. explaining its marginal $E_F$ dependence.

In addition to the QNM introduced in the main paper, there are other QNMs of this metasurface that are very far from the real operation frequency. If we extend the imaginary frequency in Fig. 4a, there is another $QNM_2$ below the $QNM_1$ introduced in the main paper as shown in the Fig. S12a. Figure S12c-d decomposes the complex diffraction coefficients of this metasurface into the contributions of the two resonant QNMs and a 'new' background response. The new background response is the sum of the contributions from QNMs that are further away from the real operation frequency than $QNM_2$ [17]. $QNM_2$ is a much more lossy resonance than $QNM_1$, so its contribution (blue dotted lines) has a very broad resonance lineshape along the real frequency axis. This resonance is almost spectrally uniform, shifting slightly up and down with $E_F$. The new background response (yellow dotted lines, BG) exhibits a non-resonant and almost independent behavior with respect to $E_F$. The complex sum of the complex diffraction coefficients of $QNM_2$ and the new background response, $QNM_2$+BG, gives the green solid lines which agrees well with the behavior of the black dotted lines in the upper panel of Fig. 4b. The gray line in the Fig. S12d is the phase difference between $QNM_1$ and the $QNM_2$+BG, which agrees well with the black solid lines in the lower panel of Fig 4b. This implies that the 'background' response spectrum in Fig. 4b can be decomposed as the sum of the $QNM_2$ and the new background response spectrum, and the marginal $E_F$ dependence of the 'background' response in Fig. 4b can be attributed to the presence of $QNM_2$ inherent in it.

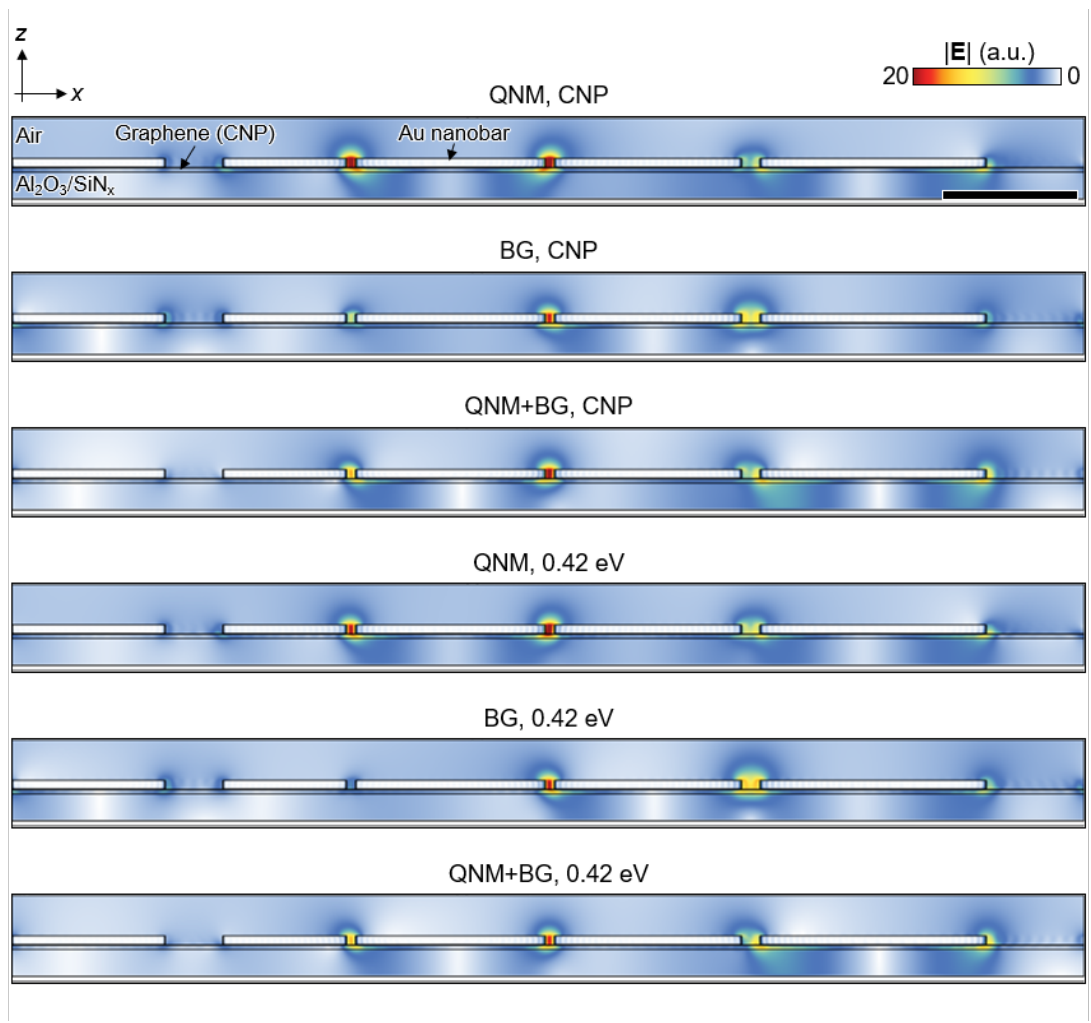

**Fig. S13.** Decomposed (QNM, BG) and reconstructed (QNM+BG) electric field intensity profiles over one grating period of the metasurface at the CNP and $E_F = 0.42$ eV.

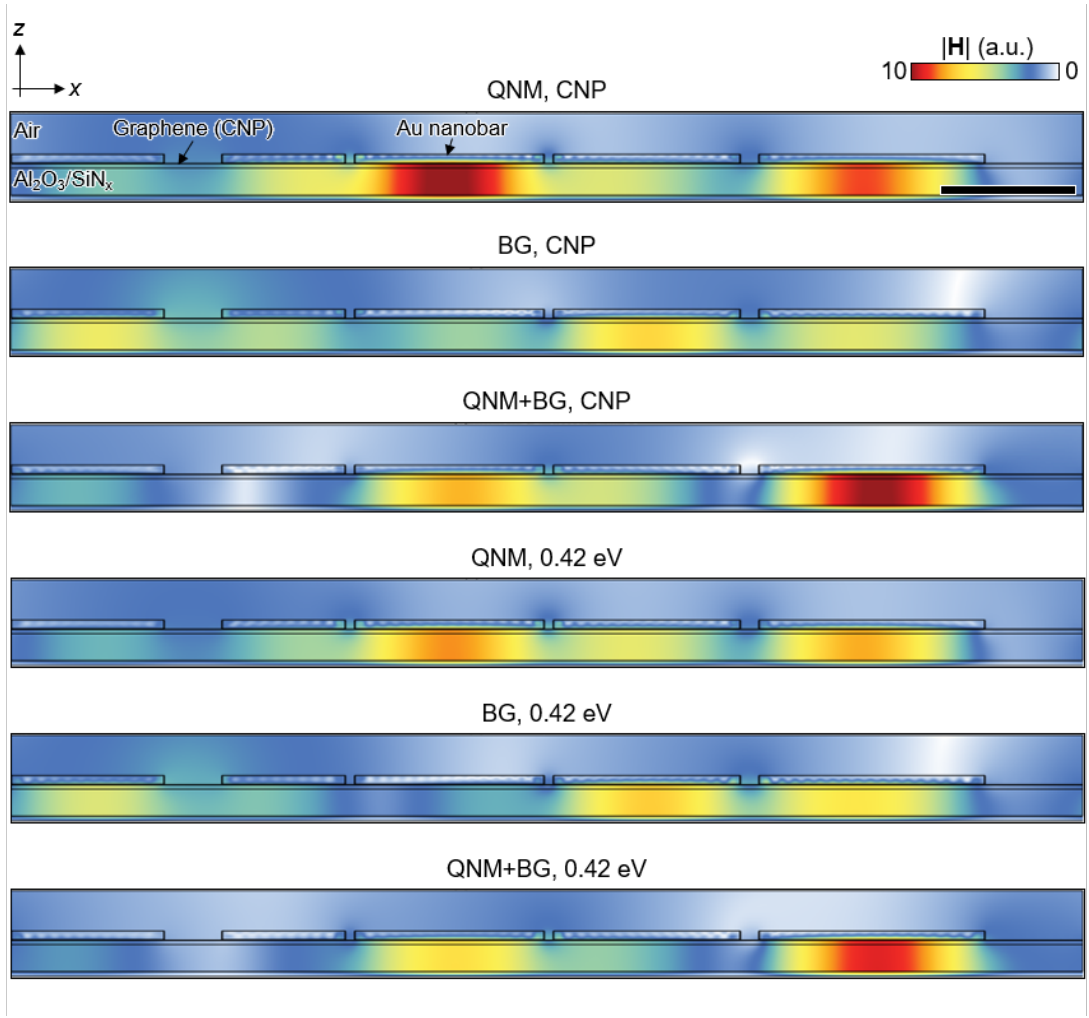

**Fig. S14.** Decomposed (QNM, BG) and reconstructed (QNM+BG) magnetic field intensity profiles over one grating period of the metasurface at the CNP and $E_F = 0.42$ eV.

## 7.3 Riesz projection and contour integrals

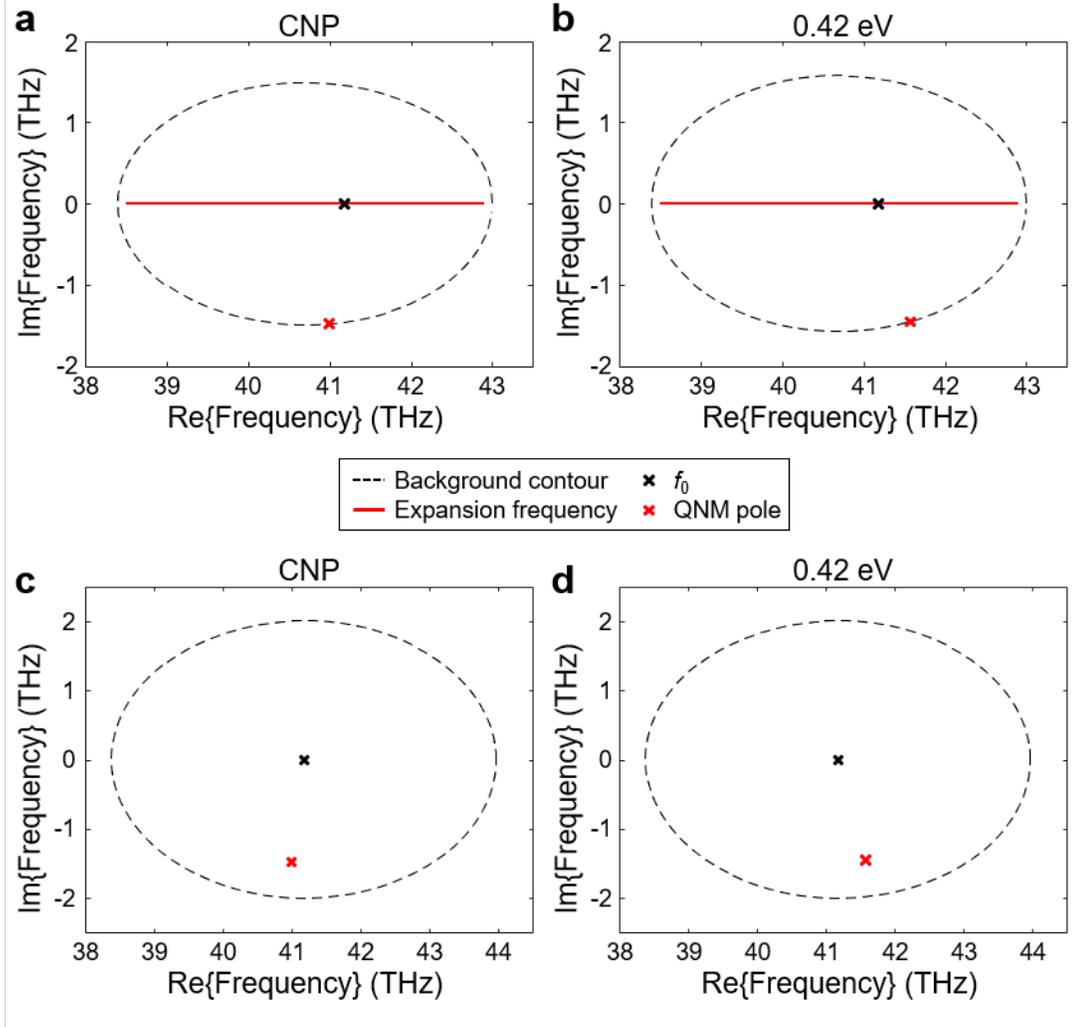

**Fig. S15. Cauchy's integral contours.** Cauchy's integral contours ($C_{BG}$, black dashed lines) to perform the Riesz projection (Fig. 4b) at **a** the CNP and **b** $E_F$ = 0.42 eV, and to decompose electromagnetic field profiles (Figs. S13-14) at **c** the CNP and **d** $E_F$ = 0.42 eV. In **c** and **d**, contour $C_k$ is a circle with a radius of 0.01 THz centered at $f_k$ (red x-marks).

To decompose complex diffraction coefficients or electromagnetic profiles, Riesz projection is performed. We firstly define $q(\omega_0)$ as the analytic continuation of the physical observable of interest. Applying Cauchy's residue theorem to $q(\omega_0)$, we obtain the expansion shown below[17]:

$$q(\omega_0) = \sum_{k=1}^{n} q_k(\omega_0) + q_{BG}(\omega_0). \tag{S10}$$

Where $q_{BG}(\omega_0)$ corresponds to the contribution of QNMs outside the region of interest, which is the non-resonant background response. The $q_k(\omega_0)$ is the Riesz projection of $q(\omega_0)$, for a particular $k$-th pole. $q_k(\omega_0)$ and $q_{BG}(\omega_0)$ are defined by the following contour integrals:

$$q_k(\omega_0) = -\frac{1}{2\pi i}\oint_{C_k}\frac{q(\omega)}{\omega-\omega_0}d\omega = \frac{1}{2\pi i}\oint_{C_{BG}}\frac{q(\omega)}{\omega-\omega_0}\prod_{\substack{m=1\\m\neq k}}^{n}\frac{\omega-\omega_m}{\omega_k-\omega_m}d\omega, \qquad (S11)$$

$$q_{BG}(\omega_0) = \frac{1}{2\pi i}\oint_{C_{BG}}\frac{q(\omega)}{\omega-\omega_0}d\omega. \qquad (S12)$$

Here, the contour $C_{BG}$ includes both the frequency of interest $\omega_0 = 2\pi f_0$ and the poles $\omega_k = 2\pi f_k$, and the contour $C_k$ includes only a single $k$-th pole. If the position of the pole is precisely defined, then using the algorithm proposed in Ref. 18, $q_k$ can be computed using only $C_{BG}$ instead of the contour integral over $C_k$. To apply this algorithm, we utilize RPExpand, an open MATLAB library[17]. To compute the Riesz projection for complex diffraction coefficients, the integration contours $C_{BG}$ are set as shown in Fig. S15a-b. The contours are set to include the real frequency range $f = [38.5, 42.9]$ THz (red solid lines) and the frequency of interest $f_0$ (black x-mark), as well as the QNM resonance frequency $f_k$ (red x-mark), which varies with $E_F$. To decompose the 0th and the -1st order diffraction coefficients, the integrands are calculated at 150 and 120 discrete points along the corresponding contour line $C_{BG}$, respectively. To calculate the electromagnetic field intensity $\mathbf{E}(\mathbf{r},f_0)$ and $\mathbf{H}(\mathbf{r},f_0)$ in Figs. S13-14, $C_{BG}$ is set as shown in Fig. S15c-d, and $C_k$ is set as a circle with a radius of 0.01 THz centered at $f_k$. To decompose the electromagnetic field intensity, integrands are calculated at 150 and 25 discrete points on $C_{BG}$ and $C_k$, respectively. Here, each integrand value is calculated using $S^4$, an open Python RCWA library, and the number of the Fourier orders utilized in the RCWA simulation is 175. All material permittivities used in the simulation are analytically continued to the complex frequency as described in the Supplementary Note 12.

## Supplementary Note 8. Optimization of a metasurface with relaxed constraints

### 8.1 2-level beam switching

As relaxed constraints, the graphene carrier mobility is increased from 500 cm$^2$/V·s to 1000 cm$^2$/V·s. Additionally, the lower bounds for the gold strip width $w_i$, and gap $g_i$ are reduced to 20 nm, while the upper bound for the gold strip height $h$ is increased to 100 nm. To account for the reduced minimum feature size, RCWA simulations are performed with a Fourier order of 350. Except for these adjustments, the optimization process follows the same methodology as described in Supplementary Note 3, including the restriction condition in Eq. S6-7. The final optimal structure is derived with the parameters $w_i$ = (1110, 1449, 22, 1096, 1194) nm, $g_i$ = (483, 312, 386, 26, 20) nm, $h$ = 41 nm, and $f_0$ = 44.16 THz.

### 8.2 3-level beam switching

The same relaxed constraints used in the optimization of the 2-level beam switching metasurface are applied. However, for the design of the 3-level beam switching metasurface, which utilizes a 0° incidence angle and the 0th, -1st, and 1st order diffraction channels, the restriction conditions of the parameter set are modified to suppress higher diffraction orders, as described in Eq. S13-14:

$$\frac{\lambda}{P} < 1, \tag{S13}$$

$$\frac{2\lambda}{P} > 1. \tag{S14}$$

For RCWA simulations, a Fourier order of 300 is used. The final optimal structure is derived with the parameters $w_i$ = (1503, 1178, 1446, 1972, 538) nm, $g_i$ = (1073, 248, 141, 79, 179) nm, $h$ = 74 nm, and $f_0$ = 45.89 THz.

## Supplementary Note 9. Device fabrication steps

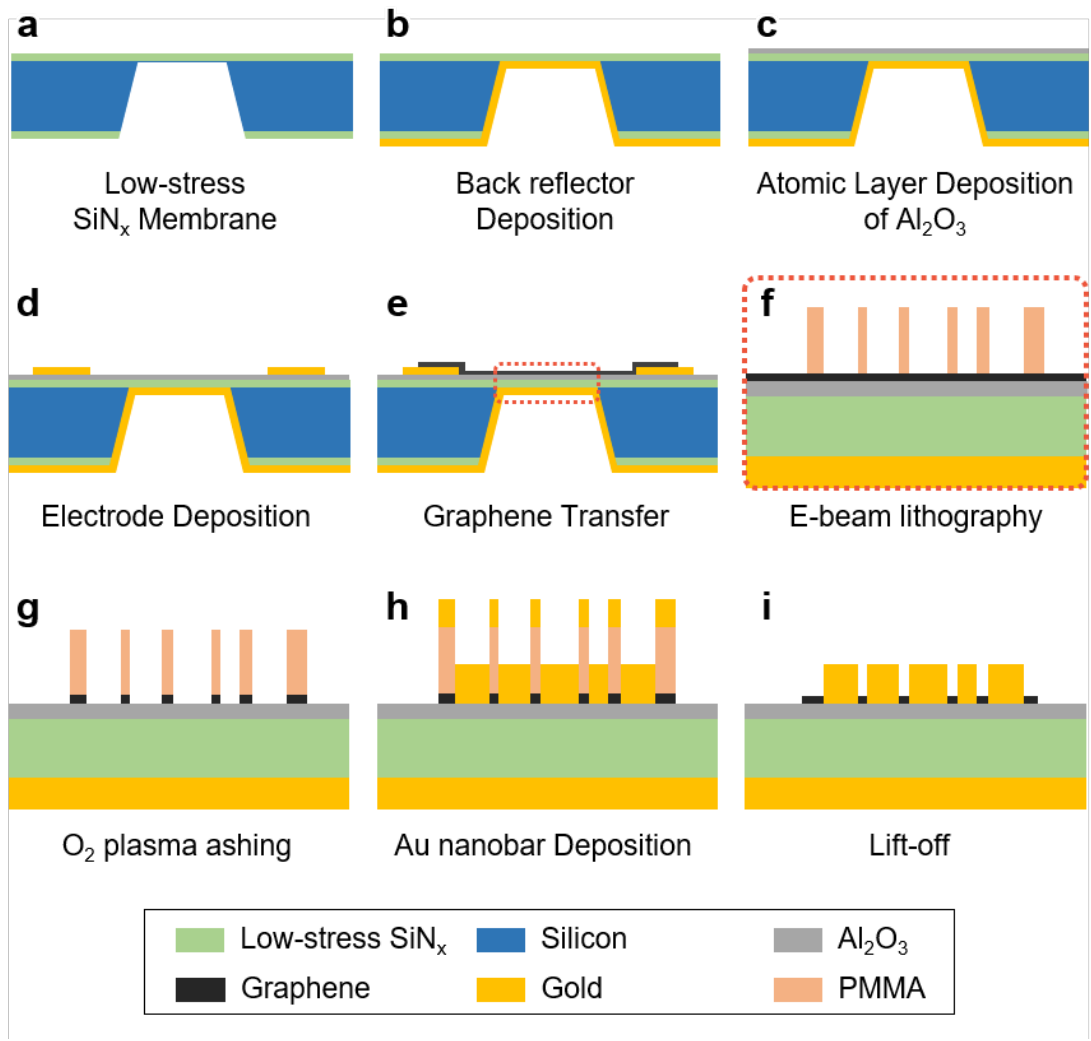

**Fig. S16. Steps for fabricating the device. a** A 200 nm-thick low-stress silicon nitride membrane supported by a silicon frame is prepared. **b** A 70 nm gold back reflector with a 3 nm Ti adhesion layer is deposited on the backside of the silicon nitride membrane using a thermal evaporator. **c** A 30 nm $Al_2O_3$ layer is deposited on the front side of the silicon nitride membrane using atomic layer deposition (ALD). **d** Electrode lines consisting of 7 nm of Ti for adhesion and 70 nm of gold are deposited on the front side of the silicon frame using photolithography and a thermal evaporator. **e** The monolayer graphene is wet-transferred to sufficiently cover the electrode lines and the entire membrane. **f** PMMA is deposited on the graphene layer and subsequently patterned on the designated area of the membrane where the metasurface will be formed using E-beam lithography (EBL). **g** The part of the graphene exposed by the patterned PMMA is etched using an $O_2$ plasma asher. **h** 6 nm of Ti for adhesion and 64 nm of gold are deposited on the exposed $Al_2O_3$ layer and patterned PMMA using a thermal evaporator. **i** The grating is formed by removing the PMMA and the gold deposited on it using a lift-off process with acetone.

## Supplementary Note 10. Reduction of the background signal fluctuation

The powermeter we utilize for our optical measurements (PM16-401, Thorlabs) has an inherent background signal fluctuation. This fluctuation appears linearly on measurement time scales of a few hundred seconds. Measuring the background optical signal in a dark-state with no incident laser light before and after the main measurement, we compensated the raw data with a linear function of time to calculate the time-compensated power $P_{compensated}$ as shown in Eq. S15 below:

$$P_{compensated}(t) = P_{raw}(t) - \left(\frac{P_{dark,end} - P_{dark,start}}{t_{end} - t_{start}}(t - t_{start}) + P_{dark,start}\right). \quad (S15)$$

Where $P_{raw}$ is the raw power measured at time $t$. In other words, it is the power value reflected from the metasurface during the gate bias cycle. $P_{dark,start}$ and $P_{dark,end}$ are the background power measured in the dark-state before and after the main measurement, respectively. $t_{start}$ and $t_{end}$ are the start and end times of the measurement, respectively.

# Supplementary Note 11. Effects of a nitrogen atmosphere

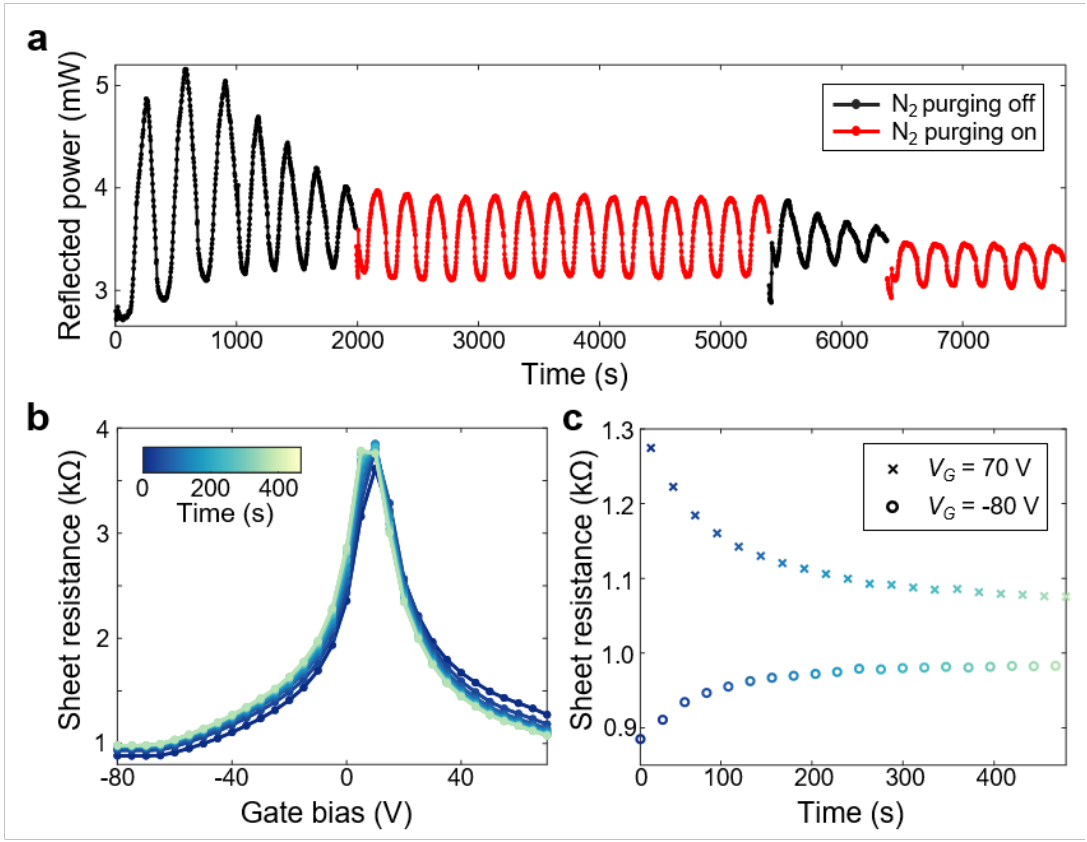

**Fig. S17. Effects of a nitrogen atmosphere around metasurfaces. a** Modulation cycles of reflected power from the metasurfaces over time. The nitrogen purging is turned on (red lines)/off (black lines) during the cycles. **b** Graphene sheet resistance-gate bias graph over time after nitrogen purging is turned on. $V_{CNP}$ is slightly shifted and saturated over time. **c** The graphene sheet resistivity over time at $V_G = 70$ V and $V_G = -80$ V, the two ends of the gate bias sweep cycle.

## 11.1 Nitrogen purging to prevent the graphene burning

During the measurement, the high power of the focused CW laser can degrade the graphene in real time. This is due to local heating caused by the high power of the laser and the high absorption of the graphene metasurface, which makes the graphene to oxidize and disappear. To remove oxygen molecules around the graphene metasurface, a nitrogen atmosphere is provided around the metasurface. Figure S17a presents the modulation range of the reflected laser power from the metasurface when nitrogen purging is on and off. Several modulation cycles are measured over time while varying the gate bias from $V_G = 50$ V to $V_G = -50$ V. When nitrogen purging is off and oxygen is present around the metasurface, the modulation range decreases rapidly after a few cycles. When nitrogen purging is on, oxygen is removed around the metasurface preventing the graphene to be burnt out, and a modulation range is maintained for many modulation cycles.

**11.2 Dirac voltage stabilization during the nitrogen purging**

When the nitrogen purging is started, the Dirac voltage of the graphene marginally shifts during the first few gate bias cycles. This is presumed to be due to the adsorbates, which are adsorbed on the graphene when it is exposed to the air, detaching during the nitrogen purging and gate bias cycles[19]. Figure S17b shows the overall shift in the sheet resistance-gate bias curve from the start of nitrogen purging until the Dirac voltage stabilizes. At both end gate biases of $V_G = -80$ V and $V_G = 70$ V (Fig. S17c), graphene sheet resistance converges to a certain value and stabilizes over the cycle. All measurements in this experiment are conducted after confirming this stabilization.

# Supplementary Note 12. Material refractive index fitting

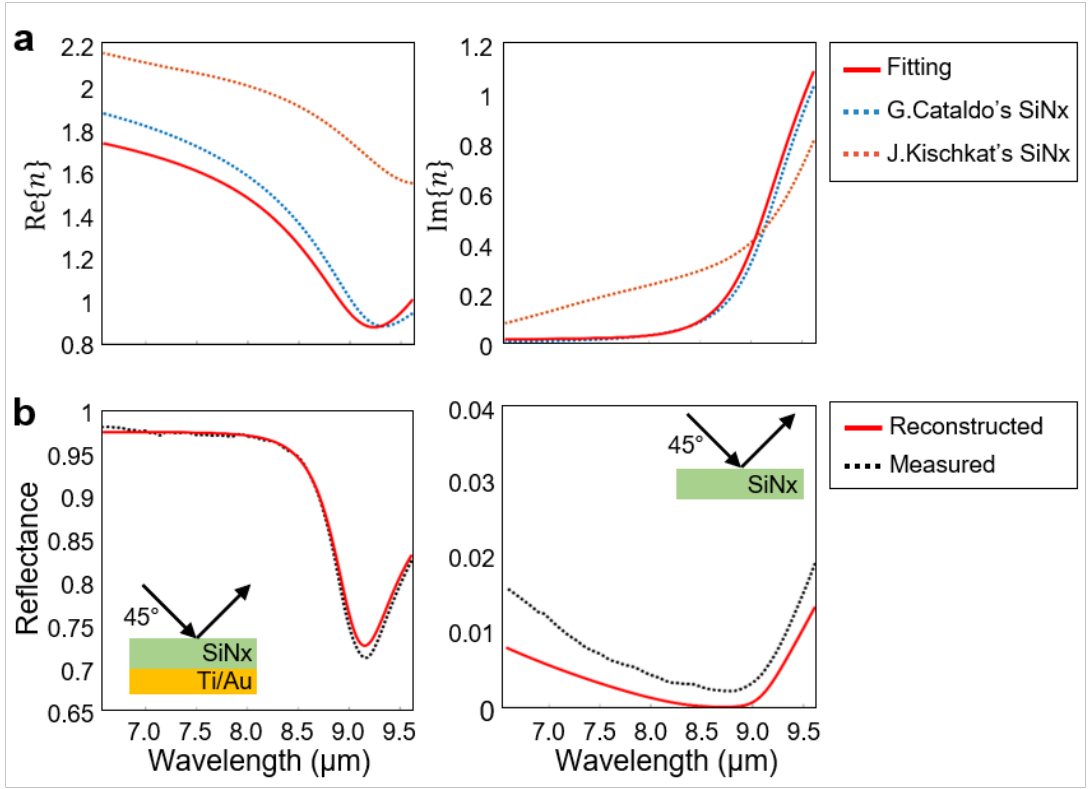

**Fig. S18. Results of fitting the refractive index of low-stress silicon nitride. a** Real part (left panel) and imaginary part (right panel) of the fitted refractive index of low-stress silicon nitride (red solid line) as a function of wavelength. For comparison, the refractive index data of silicon nitride from G. Cataldo (blue dashed line) and J. Kischkat (red dashed line) are also plotted. **b** Reconstructed (red solid line) and measured (black dashed line) reflection spectra of the silicon nitride membrane with (left panel) and without (right panel) the deposition of a back reflector.

The refractive index of the low-stress silicon nitride used in the device is determined through reflectance measurements, utilizing the same optical setup illustrated in Fig. 2a. A 200 nm-thick silicon nitride membrane is measured both with and without the deposition of a 3 nm Ti adhesion layer and a 70 nm gold back reflector. TM-polarized light at an incidence angle of 45° is used to measure reflectance over the wavelength range of 6.589 µm to 9.615 µm. The dispersions of the refractive index ($n$) and relative permittivity ($\varepsilon$) are fitted to the two measured reflectance spectra using the Brendel-Bormann model, as shown in Eqs. S16-19 [20]:

$$\varepsilon(\nu) = \varepsilon_\infty + \sum_{k=1}^{m} X_k(\nu), \quad \text{(S16)}$$

$$X_k(\nu) = \frac{i\sqrt{\pi}\,\nu_{pk}^2}{2\sqrt{2}\,a_k(\nu)\sigma_k}\left[w\left(\frac{a_k(\nu) - \nu_{0k}}{\sqrt{2}\sigma_k}\right) + w\left(\frac{a_k(\nu) + \nu_{0k}}{\sqrt{2}\sigma_k}\right)\right] \quad (S17)$$

where

$$w(x) = e^{-x^2}\left(1 + \frac{2i}{\sqrt{\pi}}\int_0^x e^{t^2}dt\right), \quad (S18)$$

$$a_k(\nu) = \sqrt{\nu^2 + i\nu_{\tau k}\nu}. \quad (S19)$$

Transfer-matrix method (TMM) is used to obtain theoretical reflectance spectra[21]. The detailed parameter values used for fitting to the Brendel-Bormann model are shown in Table S1.

| $\varepsilon_\infty$ | $\nu_{p1}$ | $\nu_{01}$ | $\nu_{\tau 1}$ | $\sigma_1$ | $\nu_{p2}$ | $\nu_{02}$ | $\nu_{\tau 2}$ | $\sigma_2$ |
|---|---|---|---|---|---|---|---|---|
| 3.753 | 1056 | 880.0 | 5.141 | 108.8 | 779.2 | 778.5 | 6715 | 1330 |

**Table S1.** Values of the Brendel-Bormann model parameters. All parameters except for $\varepsilon_\infty$ are in cm$^{-1}$.

The fitted dispersion of the refractive index of low-stress silicon nitride is shown in Fig. S18a, compared with the results obtained by G. Cataldo and J. Kischkat[22,23]. The reconstructed reflectance spectra using the fitted refractive index of low-stress silicon nitride show good agreement with the measured spectra in both cases, as shown in Fig. S18b. For the purpose of complex frequency analysis, the complex frequency is substituted into the frequency term in Eq. S16-19.

## Supplementary Note 13. Angular divergence of the reflected beam

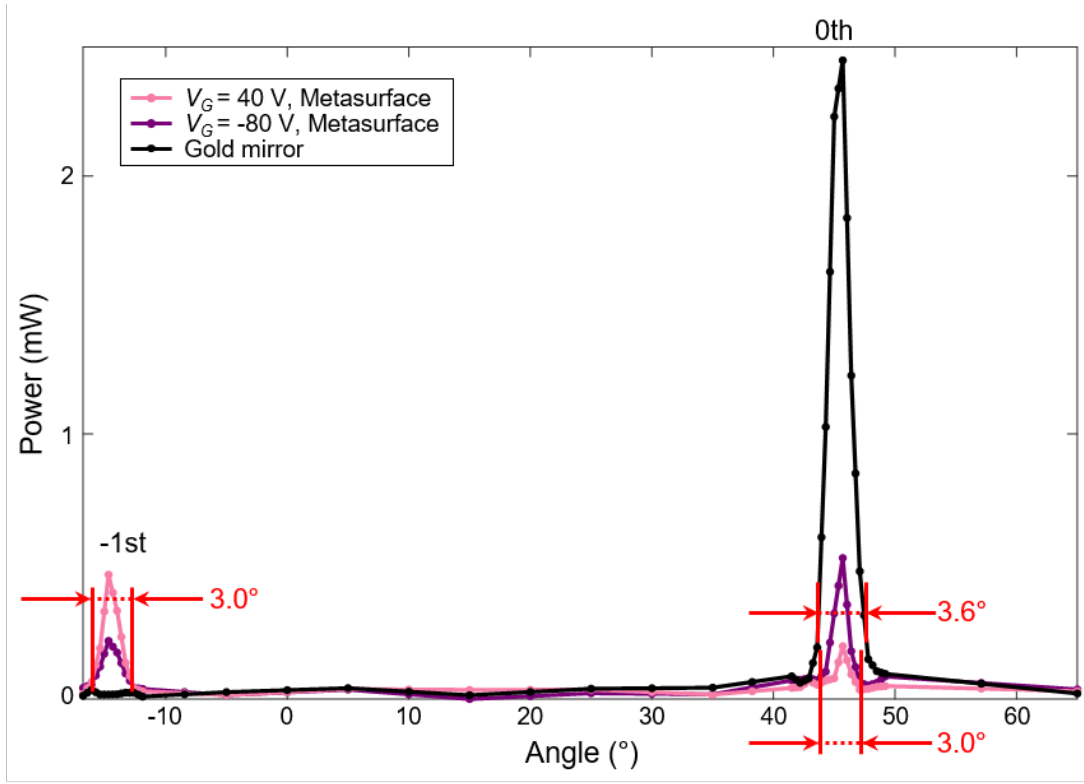

**Fig. S19.** Angular radiation pattern of the reflected beam from the gold mirror (black line) and the metasurfaces at $V_G = 40$ V (pink line) and $V_G = -80$ V (purple line). The angular divergence of main lobes of each angular radiation pattern is denoted.

To measure the angular divergence of the reflected beam of the 0th and the -1st order diffraction, the powermeter doesn't have sufficient angular resolution because it collected data in the angular range of $\Delta\theta = 7.125°$ at once. By placing a narrow slit in front of the powermeter, the angular resolution is increased to $\Delta\theta = 0.36°$. The angular divergence is measured at a frequency $f_0 = 37.05$ THz ($\lambda_0 = 8.092$ μm) with a metasurface different from the one shown in the main paper due to their different structure parameters. The size of the entire metasurface is the same as the metasurface in the main paper, and the beam spot diameter is $2w_0 = 199$ μm ($w_0$ is the beam waist.), which is smaller than the size of the entire metasurface. The angular divergence of a peak $\Theta$ is defined as the full angular width at the point where the reflected power is $1/e^2$ of the maximum reflected power of that peak, which is expressed as $\Theta = 2\lambda_0/(\pi w_0)$ approximately[24]. In our experimental setup, it is calculated to be 2.97° for an idealized Gaussian beam. Figure S19 presents the measured angular radiated power. For the specularly reflected beam from the gold mirror, the angular divergence is measured to be 3.6°. For the beam reflected from the metasurface, the main lobe presents an angular divergence of 3.0° when the gate bias is $V_G = -80$ V, and 3.0° when $V_G = 40$ V. The measured beams show marginal deviations from the ideally calculated values. This discrepancy can be attributed to spatial inhomogeneity of structure, imperfect alignment of the optical setup, and imperfect incident laser beam quality, which can slightly distort the wavefront of the reflected beam to be deviated from the ideal Gaussian beam.